\documentclass[1p]{elsarticle}
\pdfoutput=1 

\usepackage{lineno,hyperref}
\modulolinenumbers[5]
\usepackage{float}
\usepackage{rotating}
\usepackage{lscape}
\usepackage{subfig}
\usepackage{multicol}

\usepackage{xcolor}
\usepackage{soul}

\journal{Icarus}

\biboptions{authoryear}
\begin{document}

\begin{frontmatter}

\title{Venus upper atmosphere revealed by a GCM: II. Model validation with temperature and density measurements}
%% Thomas companion paper = \title{Venus upper atmosphere revealed by a GCM: II.~Detailed structure of the nightside subsidence} % of the Subsolar-Antisolar circulation

%\author{G.Gilli\fnref{myfootnote}}

\author[ia,fcul]{Gabriella Gilli}
\cortext[mycorrespondingauthor]{Corresponding author}
\ead{ggilli@oal.ul.pt}

\author[ucla,mcgill]{Thomas Navarro}
\author[lmd]{Sebastien Lebonnois}
\author[ia,fcul]{Diogo Quirino}
\author[ia,fcul]{Vasco Silva}
\author[latmos]{Aurelien Stolzenbach}
\author[latmos]{Franck Lefèvre}
\author[ucla]{Gerald Schubert}



\address[ia]{Instituto de Astrofisica e Ciências do Espaço (IA), OAL, Tapada da Ajuda, PT1349-018 Lisboa, Portugal}
\address[lmd]{Laboratoire de Meteorologie Dynamique (LMD) Paris, France}
\address[fcul]{Faculdade de Ciências, Campo Grande, PT1749-016 Lisboa, Portugal}
\address[latmos]{LATMOS, CNRS, Sorbonne Université, Paris, France}
\address[ucla]{University of California, Los Angeles, CA, USA}
\address[mcgill]{McGill University, Montreal, Canada}

\begin{abstract}
%\textcolor{red}{Thomas' changes are in red}\\
%\textcolor{blue}{Gabriella's comments to all authors are in blue}\\
An improved high resolution (96 longitude by 96 latitude points) ground-to-thermosphere version of the Institut Pierre-Simon Laplace (IPSL) Venus General Circulation Model (VGCM), including non-orographic gravity waves (GW) parameterization and fine-tuned non-LTE parameters, is presented here.
We focus on the validation of the model built from a collection of data mostly from Venus Express (2006-2014) experiments and coordinated  ground-based telescope campaigns, in the upper mesosphere/lower thermosphere of Venus (80-150~km).
These simulations result in an overall better agreement with temperature observations above 90~km, compared with previous versions of the VGCM.
Density of CO$_2$ and light species, such as CO and O, are also comparable with observations in terms of trend and order of magnitude. Systematic biases in the temperature structure are found between 80 and 100~km approximately (e.g. GCM is 20 to 40~K warmer than measurements) and above 130~km at the terminator (e.g. GCM is up to 50~K colder than observed). Possible candidates for those discrepancies are the uncertainties on the collisional rate coefficients used in the non-LTE parameterization (above 130 km), and assumptions on the CO$_2$ mixing ratio made for stellar/solar occultation retrievals. Diurnal and latitudinal distribution of dynamical tracers (i.e. CO and O) are also analyzed, in a region poorly constrained by wind measurements and characterized by high variability over daily to weekly timescale. Overall, our simulations indicate that a weak westward retrograde  wind is present in the mesosphere, up to about 120 km, producing the CO bulge displacement toward 2h-3h in the morning, instead of piling up at the anti-solar point, as for an idealized sub-solar to anti-solar circulation. This retrograde imbalance is suggested to be produced by perturbations of a $\sim$ 5 days Kelvin wave impacting the mesosphere up to 110~km (described in the companion paper \citet{Navarro2021}), combined with GW westward acceleration in the lower thermosphere, mostly above 110~km.
On the whole, these model developments point to the importance of the inclusion of the lower atmosphere, higher resolution and finely tuned parameterizations in  GCM of the Venusian upper atmosphere, in order to shed light on existing observations.

\end{abstract}


\end{frontmatter}

%\linenumbers

\section{Introduction}
Our understanding of the Venusian climate has noticeably increased with progress with General Circulation Models (GCM), powerful 3D tools to investigate the amount of data acquired by space missions, particularly in recent time by Venus Express (VEx) \citep[e.g][]{Drossart2007Nature,Piccioni2007} and Akatsuki \citep[e.g][]{Satoh2017,Fukuhara:2017} missions, as well as on-going ground-based telescope campaigns. For instance, some model experiments agree on the role of thermal tides in the vertical transport of angular momentum in the equatorial region and in maintaining the superrotation \citep{Yamamoto2006, TakagiMatsuda2007,Lebonnois2010, Mendonca2016, Yamamoto2021}. Others succeeded to reproduce the polar atmospheric structure and the  ``cold collar" \citep{Ando2016, GarateLopez2018}, in close agreement with the observations, in particular when the latitudinal variations of the cloud structure are taken into account.

Despite numerous and increasing observations, our view of the upper mesosphere and lower thermosphere (UMLT) of Venus (i.e. 80 km-140~km approximately) still remains incomplete. Therefore, GCMs are also used to understand the atmospheric conditions of high altitude regions where some key observations are lacking, especially on the dayside \citep{Brecht2012a, Gilli2017}.
%The upper mesosphere and lower thermosphere (UMLT) of Venus (i.e. 80 km-140~km approximately) is 
The UMLT is characterized by strong diurnal variations in atmospheric temperatures which drive Subsolar-to-Antisolar (SS-AS) transport of photolyzed products such as CO, O, NO, H, O$_2$ singlet delta, toward strong nighttime enhancements in the thermosphere \citep{Clancy2012}. 

VEx observations (2006-2014) revealed an even more variable atmosphere than expected: in particular, the so-called “transition” region ($\sim$ 80-120~km), between the retrograde superrotating zonal (RSZ) flow  and the day-to-night circulation, showed latitude and day-to-day variations of temperature up to 80~K above 100~km at the terminator, and apparent zonal wind velocities measured around 96~km on the Venus nighttime highly changing in space and time. Current 3D models do not fully explain those variations, and specific processes (e.g. gravity wave propagation, thermal tides, large scale planetary waves) responsible for driving them are still under investigation. 

However, it is difficult to simulate all individual wave sources of temporal and spatial variability observed within the Venus atmosphere due to resolution constraints. It is useful for a modeling approach to have a more extensive collection of measurements to provide statistical means. 
%%%%%%%%%%%%%%%%%%%%%%%%%%%%%%%%%%%%%%%%%%%%%%%%%%%
%
%  Temperature and Density: what we know so far
%
%%%%%%%%%%%%%%%%%%%%%%%%%%%%%%%%%%%%%%%%%%%%%%%%%%
% Limaye from discussion
An example of a comprehensive data inter-comparison is the work presented in \cite{Limaye2017} (hereinafter Limaye17), showing the state-of-the-art temperature and total density measurements taken during VEx lifetime by instruments onboard and by ground-based telescopes. The study by Limaye17 put in evidence that, although there is a generally good agreement between the various experiments, differences between individual experiments are more significant than measurement uncertainties, especially above 100 km, revealing a considerable temperature variability. These experiments observed the Venus atmosphere's vertical and latitudinal temperature structure, unveiling features that appear to be systematically present, such as a succession of warm and cold layers.
%%%%% ADD SOMETHING ABOUT THE CO2 DENSITY ?
\\Regarding abundances of tracers in the UMLT, \citet{Vandaele2016} reported retrievals of CO spanning the 65-150 km altitude range during the whole VEx mission with the instrument SOIR (Solar Occultation in the InfraRed) together with an updated comparison of CO measurements from the literature. 
Within the high variability observed in the short term, reaching one order of magnitude on less than one month period, they also found a systematic latitudinal trend: CO abundances in the equatorial region were larger than high-latitude (60$^\circ$-80$^\circ$) and polar (80$^\circ$-90$^\circ$) abundances by a factor 2 and 3 respectively at an altitude of 130 km. However, below 100 km, the trend is reversed (i.e., equator-to-pole increasing), as reported by previous observations \citep{Irwin2008a, Marcq2015}. This trend reversion supports the existence of a Hadley cell circulation type \citep{Taylor1995, TaylorGrinspoon2009,Tsang2008, Marcq2015}. Interestingly, the density profiles measured by SOIR on board VEx showed a distinctive change of slope with altitude occurring in the region usually lying between 100-110 km altitude, also present in the Venus International Reference Atmosphere (VIRA) model, but at a lower altitude \citep{Vandaele2016}. In VIRA, this may be caused by different observations feeding the model for CO: above 100 km, VIRA data is based on measurements reported by \citet{Keating1985}, below 100 km the values are those reported by \citet{vonZahn1983} and later updated by \citet{deBergh2006} and \citet{Zasova2006}. Simultaneous CO and temperature measurements in the lower thermosphere \citep{Clancy2012}  suggested instead that a distinctive circulation pattern would force large-scale air masses downward, creating dynamic increases in temperature and CO mixing ratio. 
SOIR evening abundances are also systematically higher than morning values above 105 km altitude, but the reverse is observed at lower altitudes.

%%%%%%%%%%%%%%%%%%%%%%%%%%%%%%%%%%%%%%%%%%%%%%%%%%%%%%%%%%%%%%%%%%%
%%%%%%%% DYNAMICS TRANSITION REGION; WHAT WE KNOW SO FAR. 
% see paper Moullet et al. 2012, Clancy et al 2012, Lellouch et al 2008, Bougher 2006, Brecht 2012
%
%%%%%%%%%%%%%%%%%%%%%%%%%%%%%%%%%%%%%%%%%%%%%%%%%%%%%%%%%%%%%%%%%%%%%%%%%%%%%%

Regarding the wind, the most relevant constraints in the mesosphere/thermosphere are provided by the diurnal/meridional distribution of CO and of the O$_2$ 1.27 $\mu$m IR nightglow, whose peak emission occurs at 96 $\pm$ 1 km in limb viewing \citep{Drossart2007}. 
Both CO and O are produced by CO$_2$ photo-dissociation during daytime and then transported to the nightside by the SS-AS circulation. 
Light species like CO are expected to pile up at the converging stagnation point of the wind field (i.e., the point where the horizontal velocity converges to zero). This stagnation point is at the anti-solar point for idealized SS-AS flow and displaced toward the morning terminator when a westward retrograde zonal flow is added. The position of the maximum and its magnitude depend on the relative values of equatorial velocity and a maximum cross-terminator velocity. \cite{Lellouch2008} observations showed a moderate (factor 2) CO enhancement near the morning terminator. To first order, this picture should be valid for the O$_2$ airglow. However, the situation is somewhat more complex in this case because the brightness of the O$_2$ emission also depends on chemistry and vertical transport (see companion paper by \citet{Navarro2021}, hereinafter Navarro21). 
\\In addition, a combination of nightside and terminator Doppler CO wind measurements took advantage of a rare solar transit in June 2012 to address several important questions regarding Venus's upper atmospheric dynamics. \citet{Moullet2012} found a substantial nightside retrograde zonal wind field near the equator, also detected in other nightside distribution studies and less pronounced in the dayside lower thermospheric circulation \citep{ClancyMuhleman1991,Gurwell1995, Lellouch2008,Clancy2012}. This was interpreted as the presence of weaker morning versus evening cross-terminator winds, likely to be associated with morning-evening asymmetric momentum drag due to enhanced gravity wave (GW) absorption over the morning thermospheric region \citep{Alexander1992, Bougher2006,Clancy2015}.
Model simulations in \citet{Hoshino2013} showed that gravity wave forcing, coupled with a 40 m/s westward retrograde zonal wind field at 80 km altitude, is sufficient to impart a strong morning-evening asymmetry, as observed apart from the observed extreme temporal variability. \citet{Clancy2015} also measured substantially weaker morning terminator winds compared with evening terminator, consistent with \citet{Hoshino2013} results. However, the theoretical basis for forcing this asymmetry and associated retrograde zonal flow is not provided yet. Overall, wind distributions during nightside show much larger changes (50-100~m/s) over timescales of days to weeks than dayside winds \citep{Clancy2015}.

%%%%%%%%%%%%%%%%%%%%%%%%%%%%%%%%%
%%%%   Open questions 

Despite observational and modeling efforts in the last decades, Venusian upper atmosphere's temperature structure and dynamical behavior still raise questions: what are the sources of the variability observed in the airglow emissions on the nightside? Solar flux variation is expected to play only a minor role in the rapid changes observed in the nightglow morphological distribution \citep{Gerard2014, Gerard2017}. Instead, gravity waves have been suggested as a source of variability but have not been demonstrated to provide the required amplitude. In the companion article Navarro21, the state-of-the-art Venus GCM developed at the Institut Pierre-Simon Laplace (IPSL-VGCM) is used as well to investigate the variability of the upper atmosphere caused by a Kelvin wave generated in the cloud deck reaching the UMLT, and a supersonic shock-like structure wave in the thermosphere. 
In this paper, this same version of the IPSL-VGCM is validated with the most recent literature on observations of Venus UMLT atmospheric region, focusing on temperature, CO$_2$, CO and O abundances, and the effect of propagating non-orographic GW on those abundances is discussed.
%DESCRIBE THE PAPER SECTION HERE

Section~\ref{Model} contains the description of the state-of-the-art of the IPSL-VGCM. The details on the formalism adopted for the non-orographic GW parameterization, with the GW parameters setup, and the improvement of non-LTE parameterization are in Appendices A and B. In this work, the outputs of the simulations obtained with the current high-resolution version (as Navarro21) are compared with a selection of data in Section~\ref{Comparison}, while an analysis of predicted diurnal and latitudinal variations of O and CO is presented in Section~\ref{Diurnal_Lat_variation}. The results are further discussed in Section~\ref{Discussion} and the possible effects of propagating GW to the predicted abundances analyzed. Section~\ref{Conclusions} is focused on conclusive remarks and recommendations for future studies.

\section{Model description and recent improvements}
\label{Model}
The IPSL Venus GCM (IPSL-VGCM) has been used to investigate all regions of the Venusian atmosphere, as it covers the surface up to the thermosphere (150 km) \citep{Lebonnois2016,Gilli2017, GarateLopez2018,Navarro2018}. The vertical model resolution is approximately $\sim$ 2-3 km between 100 and 150 km, slightly smaller below 100 km. It includes a non-orographic GW parameterization as described in details in the \ref{GWparam}, and a photochemical module with a simplified cloud scheme, with fixed bi-modal log-normal distribution of the cloud varying with altitude. The aerosol number density is calculated for each mode and an explicit description of the H$_2$O and H$_2$SO$_4$ mixing ratios in the liquid phase is included. The mass sedimentation flux is added to tracers evolution, but the microphysical scheme is not coupled with dynamics. The evaluated mass loading in the cloud is not used by the radiative module which instead use a scheme recently updated in \cite{GarateLopez2018}. That takes into account the latitudinal variation of the cloud structure based on Venus Express observations, and lower haze heating rate yielding to a better agreement with in-situ values of wind below the cloud deck (around 45 km altitude) by the Pioneer Venus probes \citep{Schubert1983}.  Chemical abundances play a crucial role in non-LTE effects (e.g., the impact of variable atomic O in the CO$_2$ cooling rates) and EUV heating processes, which are key processes in the upper atmosphere of terrestrial planets \citep{Bougher1999}. Our model is currently the only GCM coupled with photochemistry, which enables the study of the upper atmosphere and its composition self-consistently \citep[][; Stolzenbach et al. in preparation]{Stolzenbach2016}. Here we followed the same approach as in Gilli17: the molecular weigh is calculated up to 150 km according to the chemical composition of the atmosphere, and the homopause level is evaluated by the GCM. The potential impact of a variable composition below 100 km, as revealed by N$_2$ concentration measurements between 60 and 100 km of 40 $\%$ higher than the nominal value \citep[][]{Peplowski2020} was not taken into account in this study, since we estimate that it would have a negligible impact on our results.
Note that the IPSL-VGCM considers only neutral species so far, but future developments will be addressed to extend the model up to the exobase of Venus ($\sim$ 250 km altitude) and take into account the ionospheric chemistry. Moreover, the inclusion of a stochastic non-orographic gravity wave parameterization is a plus compared to other existent thermospheric models \citep{Brecht2012a, Bougher2015} which accomplished deceleration of supersonic cross terminator winds artificially by Rayleigh friction.

\subsection{Recent improvements}
\label{rec_improv}
Compared with the previous ground-to-thermosphere VGCM described in \cite{Gilli2017} (hereinafter Gilli17), we include here several improvements in both radiative transfer code and non-LTE parameterization. The details of the improved non-LTE parameterization are provided in \ref{nlte_improv}. Also, we increased the horizontal resolution from 7.5$^\circ \times$5.625$^\circ$ to 3.75$^\circ \times$1.875$^\circ$ resulting in qualitative changes of the UMLT nighttime circulation dynamics, as explained in Navarro21. The predicted thermal structure presented in Gilli17 already captured a succession of warm and cold layers, as observed by VEx, but localized data-model discrepancies suggested model improvements. At the terminator and at nighttime thermospheric temperatures were about 40–50~K colder and up to 30~K warmer, respectively. The altitude layer of the predicted mesospheric local maximum (between 100 and 120 km) was also higher than observed, and systematically warmer, indicating an overestimation of daytime heating rates produced by absorption of IR solar radiation by CO$_2$ molecules. Among several sources of discrepancies, we identified one possible candidate. First, we discarded the effect of the uncertainty in the collisional rate coefficient K$_{v-t}$ used in the non-LTE parameterization because, as shown in Figure 12 in Gilli17, the variation associated with this coefficient only modified the temperature profiles in the thermosphere above $10^{-2}$ Pa (about 120 km), not below. We focused instead on the heating rate used as ``reference" in Gilli17, extracted from non-LTE radiative transfer line-by-line model results in \citep{Roldan2000}. In that model, the kinetic temperature and density input profiles were taken from VIRA \citep{Keating1985, Hedin1983}. Nevertheless, VEx observations revealed that VIRA is not representative of the atmosphere of Venus above 90 km \citep{Limaye2017}. VIRA is based on Pioneer Venus measurements below 100 km and above 140 km, leaving a gap where the empirical models perform poorly \citep{Bougher2015}.
New measurements of the thermal structure of the Venusian atmosphere offer the opportunity to improve these empirical models. Therefore, we decided to perform here a semi-empirical fine-tuning of the non-LTE parameterization implemented in Gilli17, based on temperature profiles by SOIR and SPICAV, as described in the next session. Note that we first performed baseline simulations with a less computationally intensive version (i.e. low horizontal resolution version 7.5$^\circ$ x 5.625$^\circ$ as in Gilli17). Successively, we increased the horizontal resolution to 3.75$^\circ$ x 1.875$^\circ$, without significant changes in the heating/cooling rates above 90 km. The combination of this high horizontal resolution and a more realistic representation of zonal winds \citep{GarateLopez2018} and temperature gave rise to a very unique GCM, able to describe the circulation of the upper atmosphere of Venus with unprecedented details, as described in the companion paper Navarro21.

%\subsection{Improved Non-LTE parameterization set-up}
%\subsubsection{Methodology}

\vspace{1cm}

\begin{table}
\centering
\begin{scriptsize}
\begin{tabular}{l c c}
\hline\hline
\textbf{Parameter description} & \textbf{Best-Fit} & \textbf{Gilli17} \\
\hline\hline
Solar Heating per Venusian day, $Vd$ [$K day^{-1}$] & 15.92  & 18.13  \\
Cloud top pressure level, $p_0$ [$Pa$] & 1985 & 1320 \\
Pressure below which non-LTE are significant, $p_1$ [$Pa$] & 0.1 & 0.008 \\
Central Pressure for LTE non-LTE transition, $ptrans$ [$Pa$] & 0.2 & 0.5 \\
\hline\hline
\end{tabular}
\end{scriptsize}
\caption{"Best-fit" values from the improved non-LTE parameterization used in this work, compared with previous version in Gilli17. See Appendix B for details.}
\label{tableNLTE}
\end{table}

\subsection{Fine-tuning of non-LTE parameters}
\label{compar_SOIR-SPICAV}
In order to fine-tune and select the set of ``best-fit" parameters listed in Table \ref{tableNLTE} we performed about 100 GCM test simulations with the goal of reducing as best as possible the discrepancies found between the averaged temperature profiles of the IPSL-VGCM and VEx dataset, in particular with SOIR/VEx \citep{Mahieux2012} and SPICAV/VEx \citep{Piccialli2014} profiles at the terminator and at night time, respectively. The major difficulty of this fine-tuning exercise was to find a combination of parameters described in \ref{nlte_improv} which also fulfills model numerical stability requirements (e.g. that allowed to run the model for several Venus days). The range of parameters that have been tested in this fine-tuning exercise is:  15.576 K to 24.60 K/day for the Solar heating per Venusian day,  1200 to 4350 Pa for the cloud-top pressure level,  6$\times 10^{-3}$ to 2.4 Pa for the pressure below which non-LTE effects are significant, and from 0.10 to 5 Pa for the center of the transition between non-LTE and LTE regime. The collisional rate coefficient was not changed compared to Gilli17. It is fixed at 3$\times 10^{-12}$, that is a "median" value commonly used in terrestrial atmosphere models.

\begin{figure}[!htbp]
    \centering
    \includegraphics[width=0.66\linewidth]{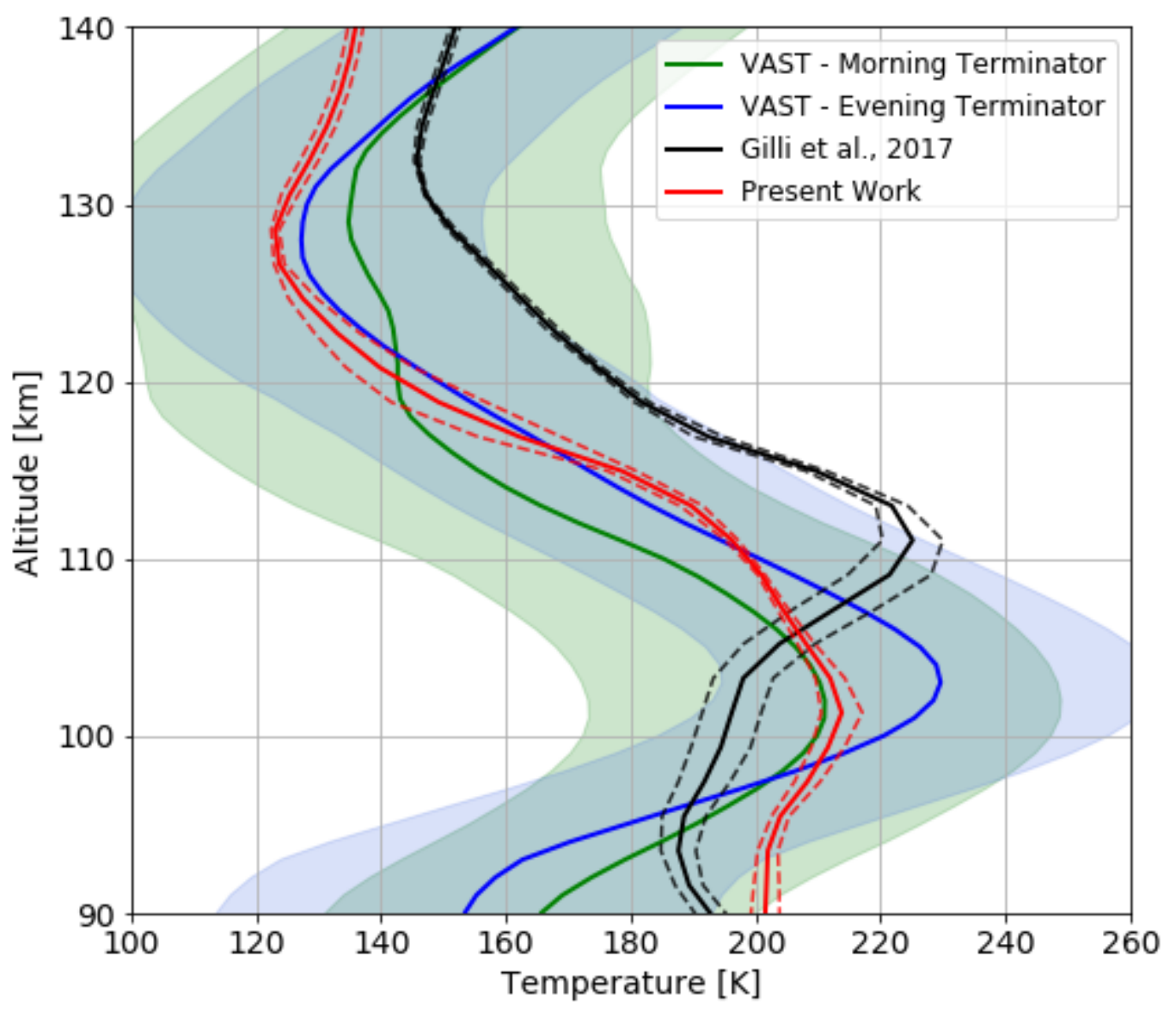}
     \includegraphics[width=0.66\linewidth]{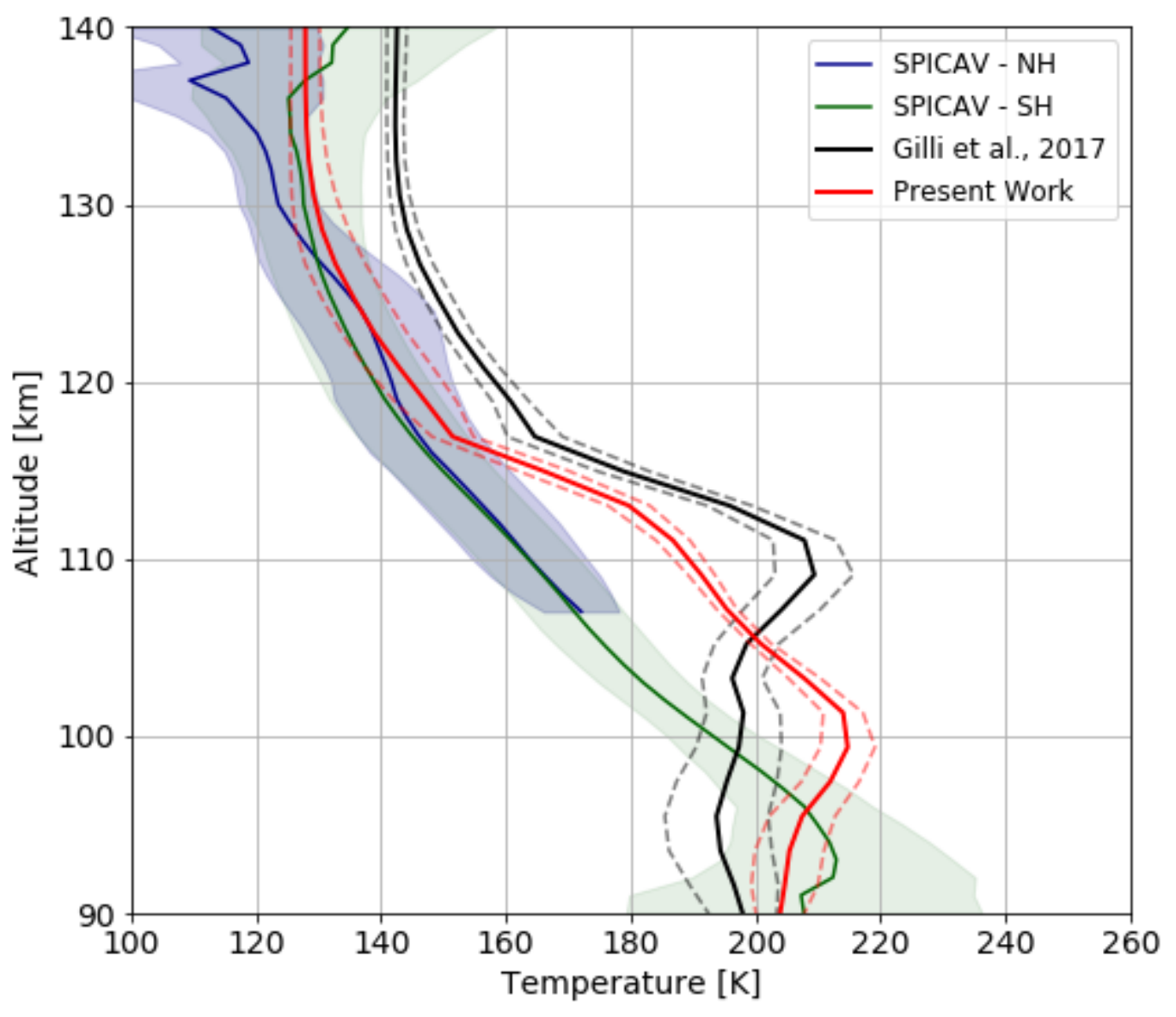}
    \caption{Averaged temperature vertical profiles as function of altitude predicted by the IPSL-VGCM before (solid black line) and after (solid red line) improving the non-LTE parameterization, together with observational data from Venus Express. Top panel: temperature retrieval results at the terminator (green for morning and blue for evening terminator) at the latitude bin 30ºN-60ºN extracted from the Venus Atmosphere from SOIR data at the Terminator (VAST) database \citep{Mahieux2015}. SOIR standard deviation is also plotted as shaded green and blue areas. Bottom panel: Temperature retrieval results from SPICAV/VEx for the Northern Hemisphere (SPICAV - NH), 30ºN-50ºN in dark blue, and for the Southern Hemisphere (SPICAV - SH), 30ºS-50ºS in green (after \cite{Piccialli2015}).
    Standard deviation of the predicted temperature mean, in the same local time and latitude range as observed, is represented by dashed lines.}
    \label{soir_spicav_vgcm}
\end{figure}

\begin{figure}[!htbp]
%    \centering
    \includegraphics[width=1.1\linewidth]{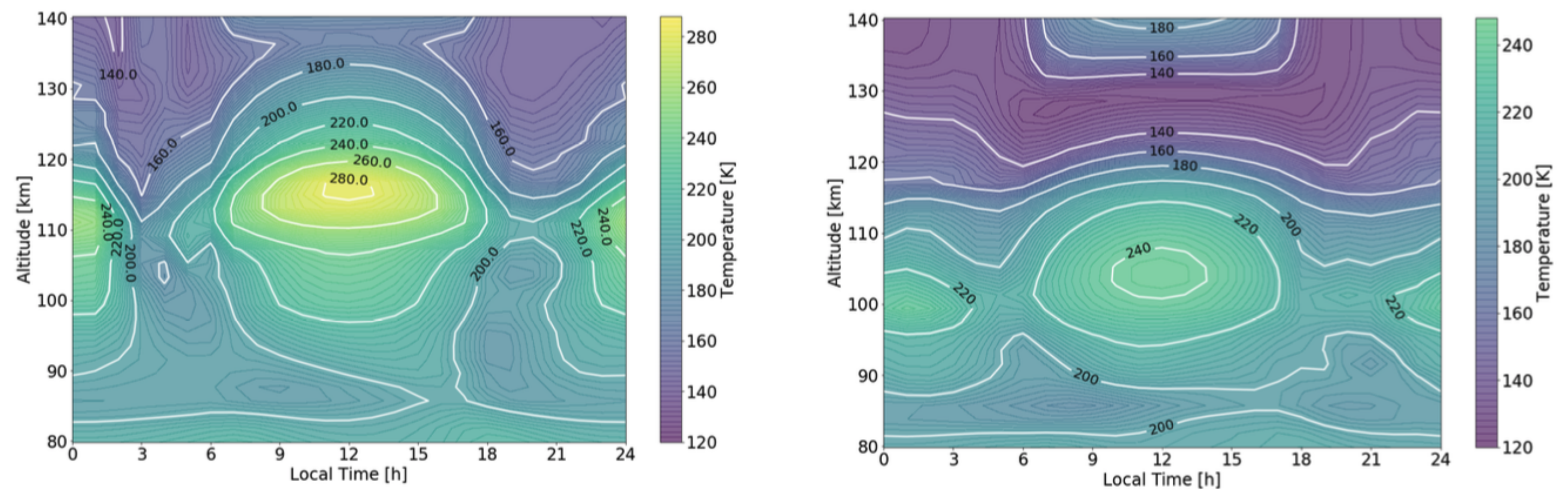}
\caption{Temperature field (local time versus altitude) obtained with IPSL-VGCM before (left panel) and after (right panel) the non-LTE parameterization improvements, listed in Table \ref{tableNLTE}. }
    \label{TEMP_maps_LCT_ALT_improvement}
\end{figure}

Figure \ref{soir_spicav_vgcm} shows an example of this improvement. Predicted upper mesospheric temperature (above 110 km approximately) are noticeably reduced by 20 K compared with the previous IPSL-VGCM version in Gilli17, and both the intensity and the altitude of the warm layer around 100~km are in better agreement with SOIR profiles. At nighttime the agreement is improved above 100 km with colder simulated temperature by 20 K, closer to observations. Above approximately 115~km, GCM results are within SPICAV error bars, though there are still data-model differences of about 20 K in the region between 100 and 115 km. The peak observed around 90-95~km is still higher in the model (around 100 km), despite a significant improvement.  
A detailed comparison of IPSL-VGCM results with a number of temperature measurements by VEx and ground-based experiments is provided in the next section.
%\textbf{FIGURE 3: Temperature field LT vs altitude: Gilli+2017 vs this work}
Temperature cross section (local time versus altitude) before and after the improvement of non-LTE parameterization is shown in Figure \ref{TEMP_maps_LCT_ALT_improvement}. The peak of temperature around midday and midnight is reduced by about 40 K and 20 K, respectively, and it is located about 10 km lower than predicted in Gilli17 (e.g. around 105 km and 100 km at midday and midnight, respectively). The temperature structure predicted by the IPSL-VGCM used in this work is in better agreement not only with VEx observations (see next section), but also with the National Center for Atmospheric Research (NCAR) Venus Thermospheric General Circulation Model presented in \cite{Brecht2012a} (see their Figure~1).

\section{IPSL-VGCM high resolution simulations: model-data validation above 90 km}
\label{Comparison}
The results discussed from this section onwards were performed using high horizontal resolution (96$\times$96$\times$78) simulations of the state-of-the-art IPSL-VGCM described in section \ref{Model}. This is the highest resolution ever used for a Venus upper atmosphere model, corresponding to grid cells of approximately 200~km in the meridional direction and 400 km in the zonal direction, at the equator. This increased resolution allows to resolve smaller-scale waves compared to those in \citet{Gilli2017} and to shed a light on the observed variability in the nightside of Venus (see Navarro21). We aim to present an exhaustive data-model comparison with the state-of-the-art IPSL-VGCM, focusing on the atmospheric layers above 85-90~km, using the most complete compilation of Venus temperature and CO$_2$ density dataset so far, obtained both from space mission and ground based observations as in Limaye2017. Moreover, a selection of CO retrieved measurements by SOIR/VEx at the terminator \citep{Vandaele2016}, by VIRTIS/VEx during daytime \citep{Gilli2015} and from Earth-based telescopes  \citep{Clancy2012} are considered here. For the atomic oxygen, only indirect O nighttime densities retrieved from O$_2$ nightglows as in \citet{Soret2012} are available for this comparison exercise. The instruments and measurements used in this validation exercise are listed in Table \ref{Table_obs}.

%%%%%. TABLE LIST OBSERVATIONS USED IN THIS WORK
%%%
%

%\begin{sidewaystable}[!htbp]
\begin{landscape}
\begin{table}
%\begin{center}
\begin{tiny}
\begin{tabular}{l |c |c |c |c |c |l}
\hline
\textbf{Experiment/Instrument} & \textbf{Method}  & \textbf{Lat Coverage} & \textbf{LT Coverage} & \textbf{Alt Coverage }& \textbf{Retrieved variable} & \textbf{References} \\
\hline
\textit{Remote sensing observations}
\\\textit{from spacecraft}
\\Magellan     & Radio occultation & N/S hemisphere & night/day side & 38-100 km& Temperature & \citet{Jenkins1994}
\\ Venera 15    & Radio occultation & N/S hemisphere & night/day side & 38-100 km & Temperature & \citet{Gubenko2008, Haus2013}
\\ Pioneer Venus & Radio occultation & N/S hemisphere & night/day side & 38-100 km & Temperature & \citet{Kliore1985, Seiff1985}
\\ VeRa/VEx     & Radio occultation & N/S hemisphere & night/day side & 38-100 km & Temperature, CO$_2$ & \citet{Tellmann2009}
\\SPICAV/VEx    & Stellar occultation  & N/S hemisphere & night side & 90-140 km & Temperature, CO$_2$ & \citet{Piccialli2010}
\\ SOIR/VEx     & Solar occultation   & N/S hemisphere & terminators & 70-170 km & Temperature, CO$_2$, CO  & \citet{Mahieux2010,Vandaele2016}
\\ VIRTIS-H/VEx Limb & non-LTE 4.7 $\mu$m CO band     & N hemisphere & day side & 100-150 km &  Temperature, CO  & \citet{Gilli2015}
\\ VIRTIS-H/VEx      & 4.3 $\mu$m CO$^2$ band       & N/S hemisphere  & night side & 65-80 km & Temperature & \citet{Migliorini2012}
\\ VIRTIS-M/VEx Nadir & 4.3 $\mu$m CO$^2$ band &  N/S hemisphere & night side  & 65-80 km & Temperature & \citet{Haus2013, Grassi2014}
\\VIRTIS-M/VEX  & O$_2$ nightglow at 1.27 $\mu$m & N/S hemisphere & night side & $\sim$ 96 km & O & \citet{Soret2012}
\\ VExADE-AER/VEX   & Aerobraking  & 70$^{\circ}$N-80$^{\circ}$N & morning terminator   &   130-140 km & CO$_2$ & \citet{Muller-Wodarg2016}
\\ VExADE-TRQ/VEX   & Torque measurements  & 70$^{\circ}$N-80$^{\circ}$N & 78-98$^{\circ}$ SZA & 165-200 km & CO$_2$ & \citet{Persson2015}
\\ HMI/SDO        & Venus transit  &  N/S hemisphere  & Terminator   & 70-110 km & Temperature & \citet{Tanga2012, Pere2016}
\\ \textit{Ground-based observations}
\\THIS/HIPWAC & non-LTE CO$_2$ emission & N/S hemisphere & day side &  110 km  &  Temperature &\citet{Sornig2008, Krause2018}
\\ JCMT        & sub/mm CO line absorption & N/S hemisphere & night/day side  & 75-120 km &  Temperature, CO & \citet{Clancy2012}
\\ HHSMT       & sub/mm CO line absorption & N/S hemisphere& night/day side & 75-110 km   & Temperature &  \citet{Rengel2008} 
\\ AAT, IRIS2, CFHT, CSHELL & O$_2$ nightglow at 1.27 $\mu$m & N/S hemisphere & night side & 85-110 km & (mean) rotational temperature & \citet{Crisp1996,Bailey2008, Ohtsuki2008} \\
\hline

\end{tabular}
\end{tiny}
%\end{scriptsize}
\caption{Observations of Temperature, CO$_2$, CO and O densities used in this paper (see Figures \ref{Night_data_VGCM}-\ref{CO2_density}). Adapted from Table 1 in \citet{Limaye2017}}
\label{Table_obs}
%\end{center}
\end{table}
\end{landscape}
%\end{sidewaystable}

\subsection{Temperature and total density}
Our current knowledge of the Venus temperature structure comes from empirical models such as VIRA \citep{Keating1985, Seiff1985} and VTS3 \citep{Hedin1983}, mostly based upon Pioneer Venus orbiter observations (PVO), and from many ground based and VEx observations \citep[e.g.][]{Clancy2008, Clancy2012, Migliorini2012, Grassi2010, Mahieux2010, Gilli2015, Piccialli2015, Krause2018}.
A comprehensive compilation of temperature retrievals since the publication of VIRA to Venus Express era was given in Limaye17, which provided for the first time a consistent picture of the temperature and density structure in the 40-180 km altitude range. This data intercomparison also aimed to make the first steps toward generating an updated VIRA model.

\begin{figure}[!htbp]
 \centering
    \includegraphics[width=0.7\linewidth]{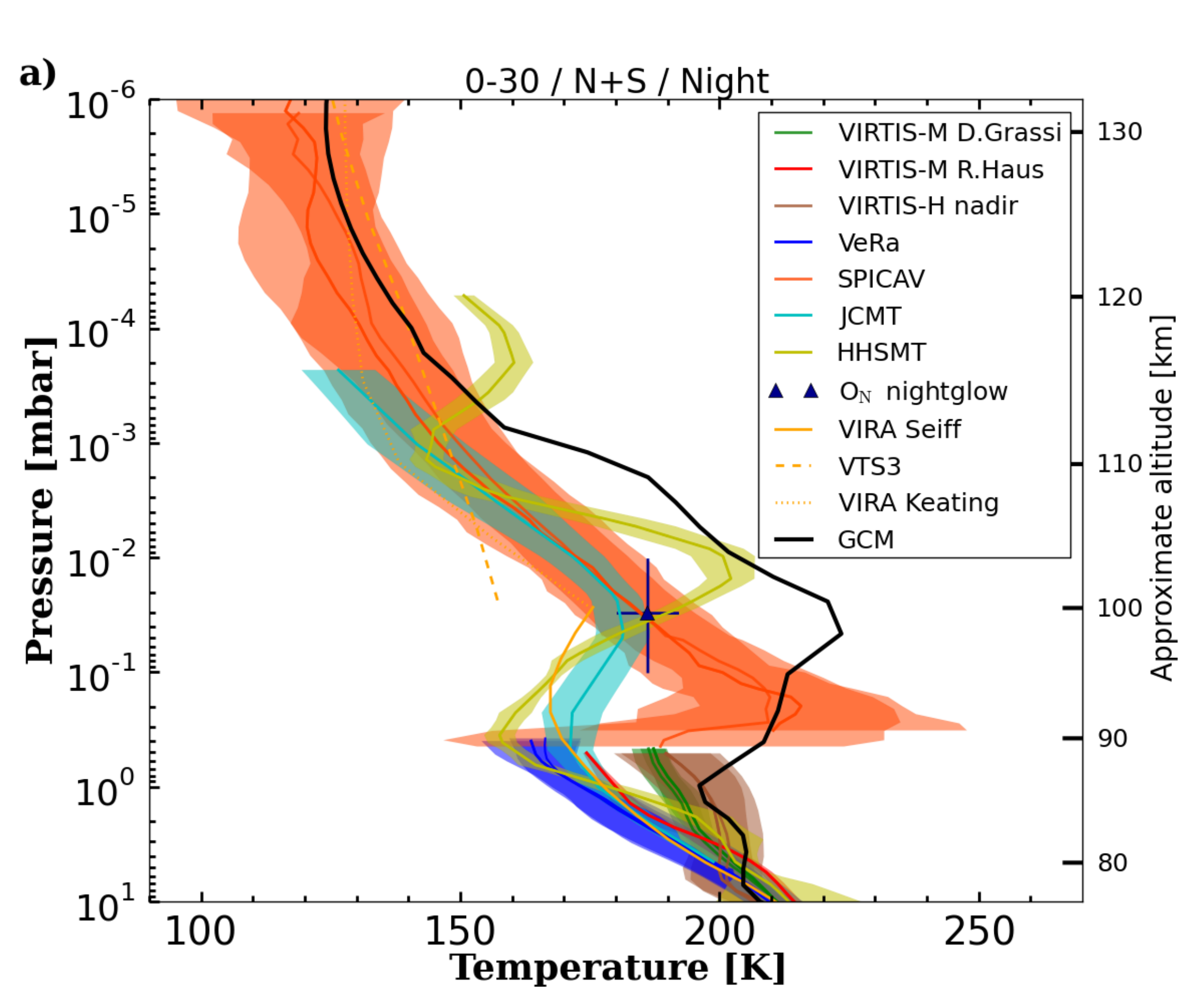}
    \includegraphics[width=0.7\linewidth]{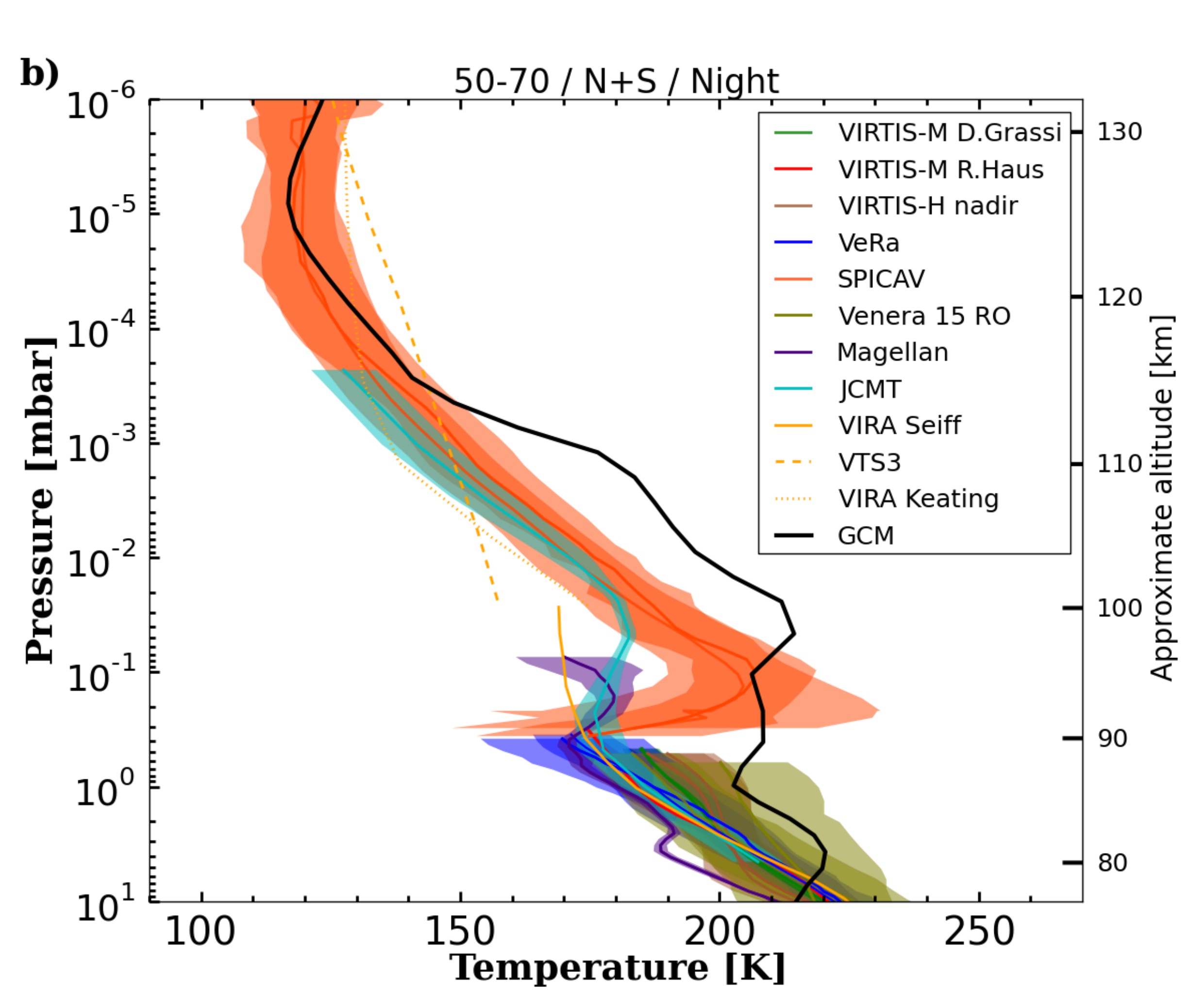}
    \caption{Compilation of available nighttime Venus temperature profiles above 80 km both from spacecrafts and ground based telescopes versus model predictions averaged at equatorial latitudes 0-30$^\circ$ from Southern and Northern hemispheres (top panel) and 50$^\circ$-70$^\circ$ N/S (bottom panel). Corresponding approximate values for altitude is given on the right hand side of the panel.
     Panel a) and panel b) are adapted from Figure 15 and 17 in \citet{Limaye2017}, respectively. See text for details. Uncertainties (one standard deviation) are either plotted as colored areas for averaged profiles in the same bin (Venus Express datasets, JCMT, HHSMT) or as error bars (O$_2$ nightglow). The IPSL-VGCM predicted profile is plotted with solid black line.}
    \label{Night_data_VGCM}
\end{figure}

\begin{figure}[!htbp]
\centering
\includegraphics[width=0.7\linewidth]{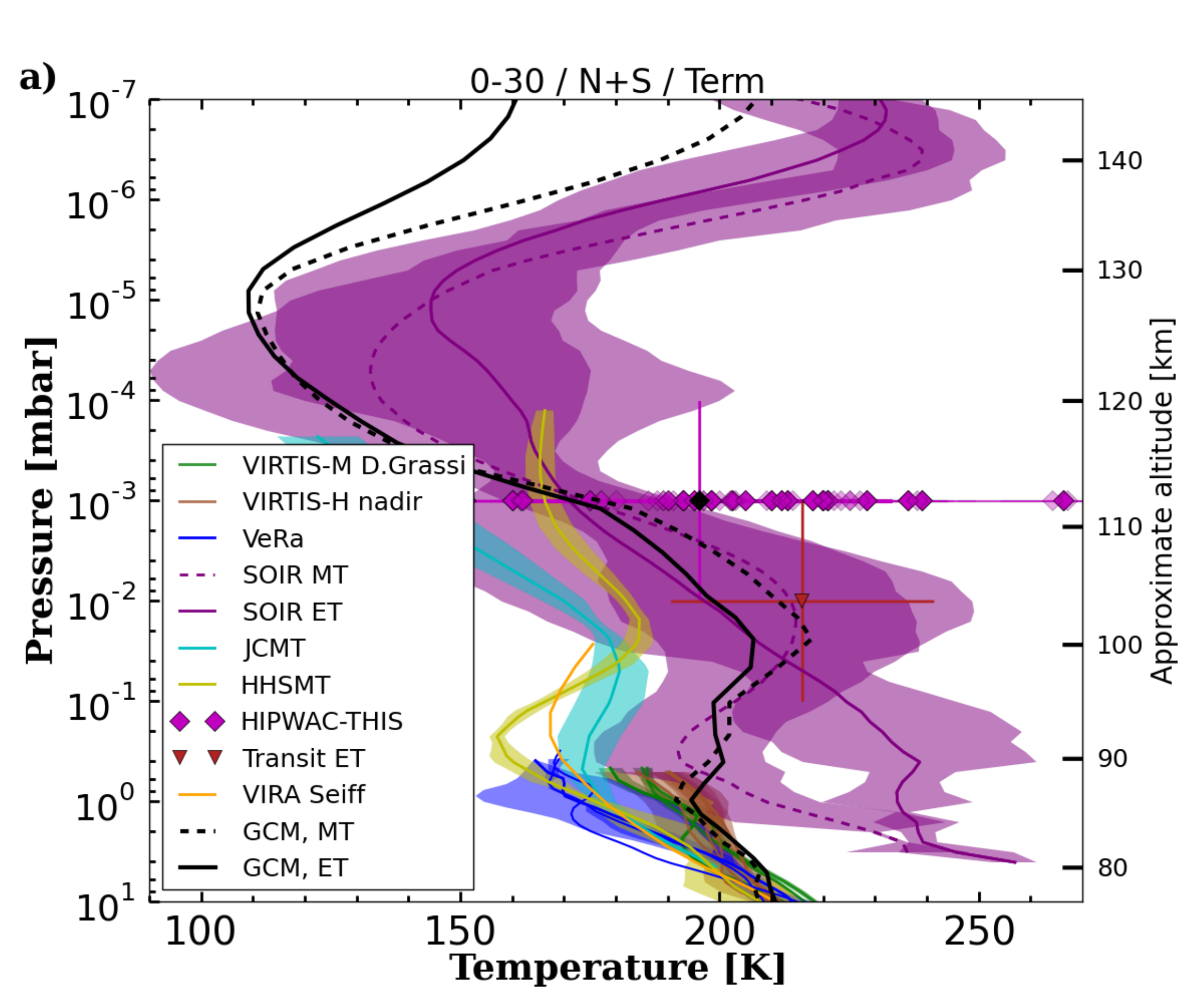}
\includegraphics[width=0.7\linewidth]{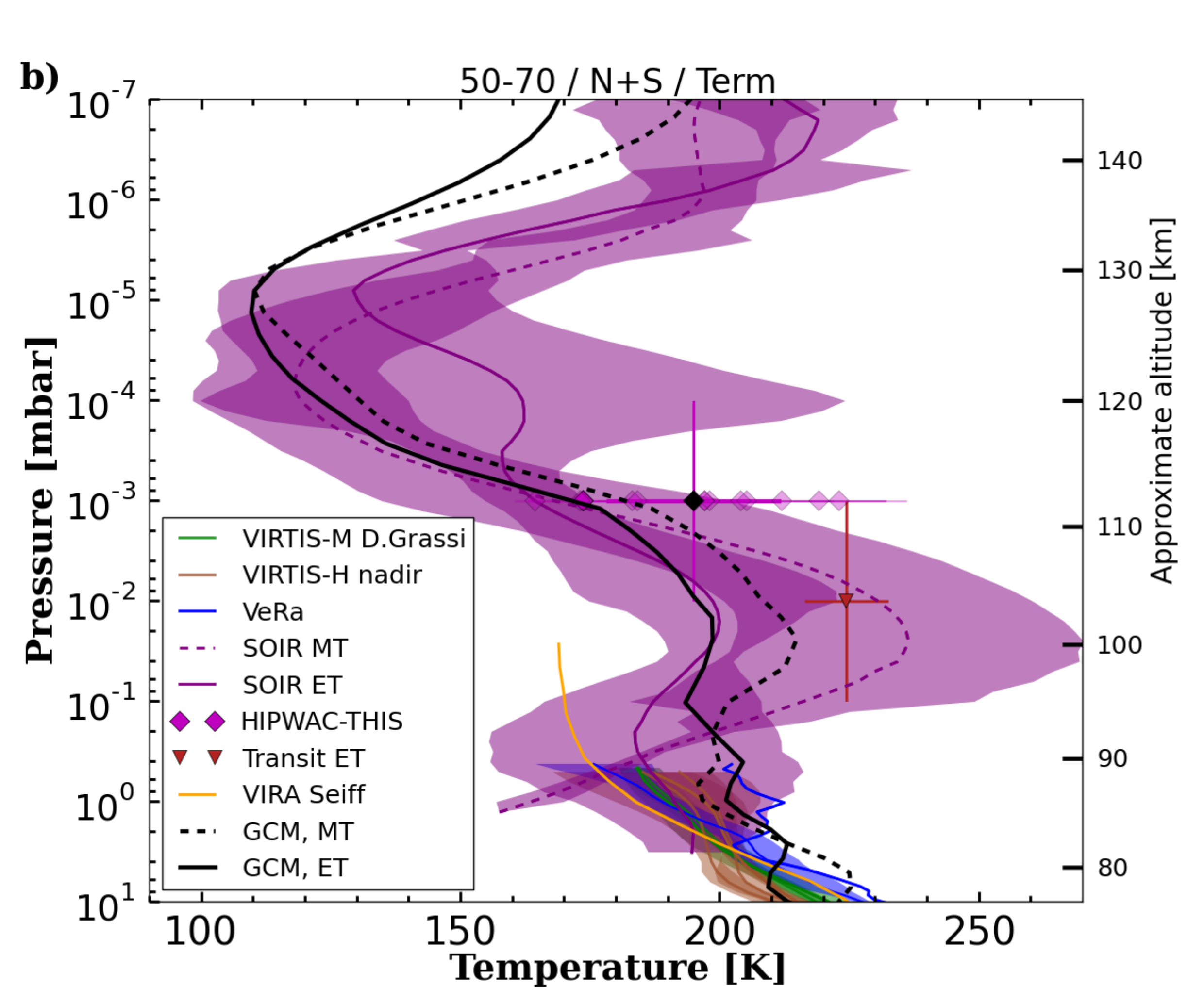}
    \caption{Same as Figure \ref{Night_data_VGCM} but at the terminators (morning and evening) and with the addition of the mean temperature values retrieved at the evening terminator during the Venus transit in 2012. See text for details. Panel a) and panel b) are adapted from Figure 15 and 17 in \citet{Limaye2017}, respectively. The IPSL-VGCM predicted profiles are in solid and dashed black lines for the ET and MT, respectively.}
    \label{Terminator_data_VGCM}
\end{figure}

\begin{figure}[!htbp]
\centering
 \includegraphics[width=0.7\linewidth]{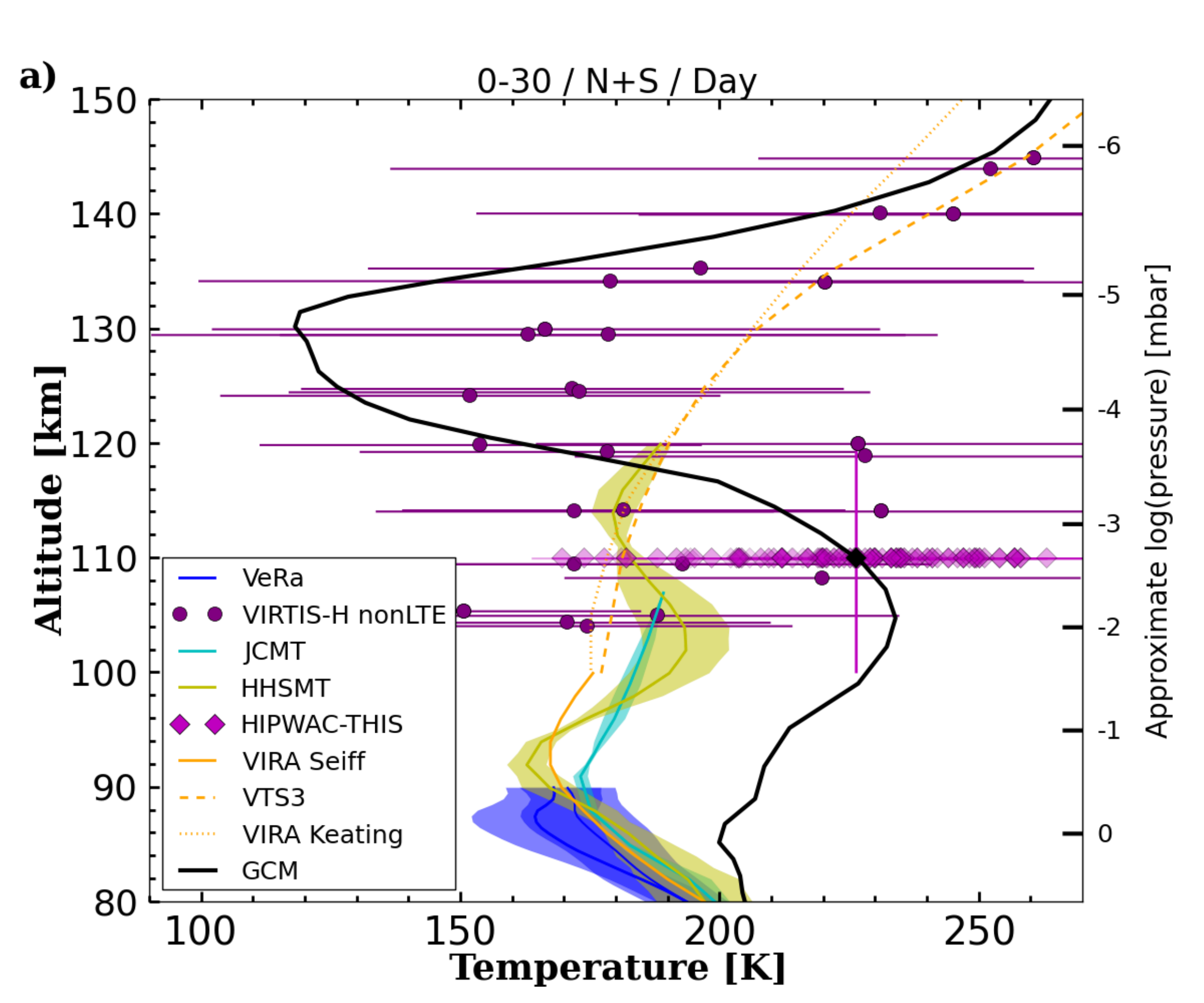}
\includegraphics[width=0.7\linewidth]{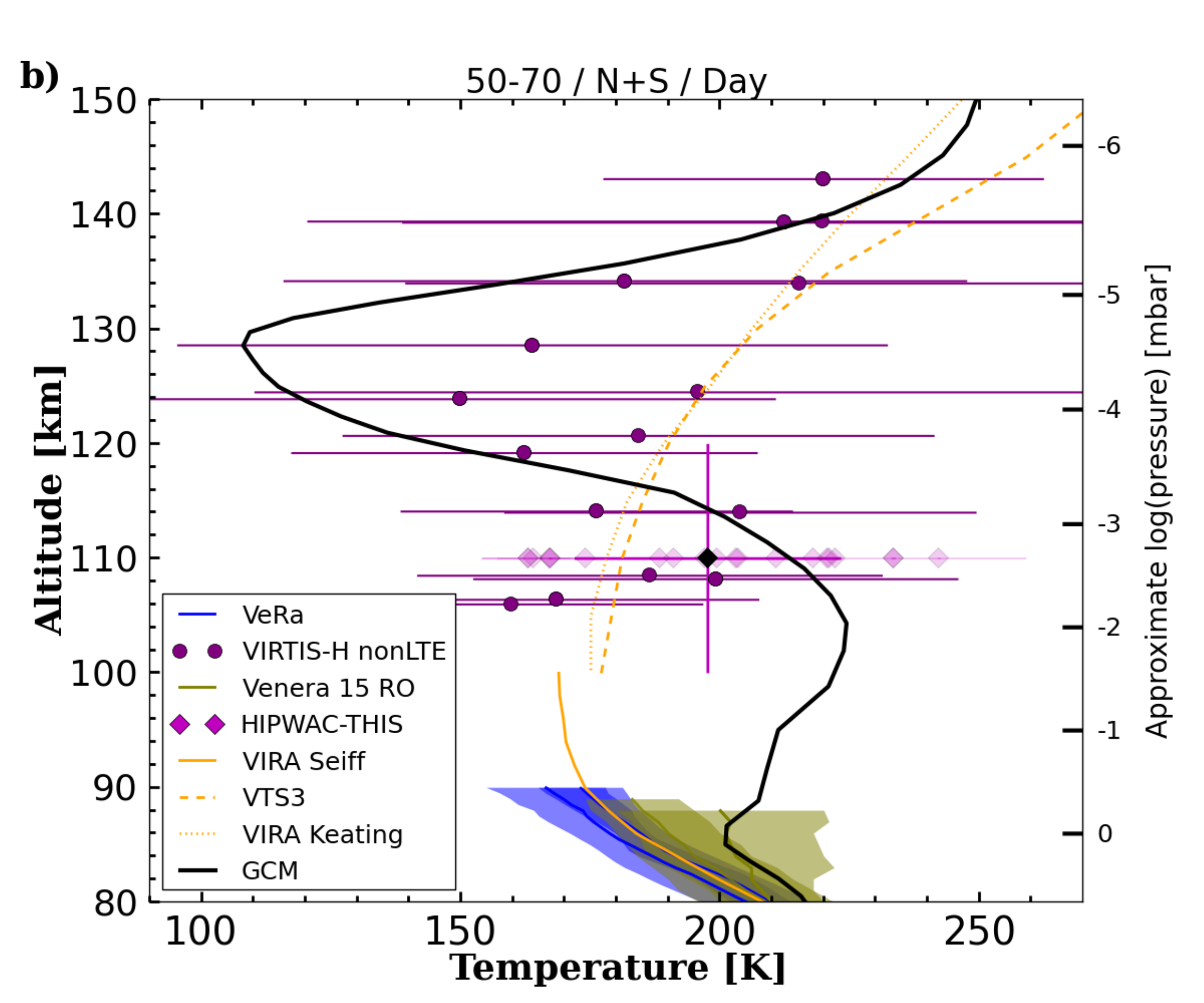}
    \caption{Same as Figure \ref{Night_data_VGCM} but for daytime and with the addition of temperature retrieved from CO non-LTE emissions observed by VIRTIS/VEx. See text for details.  Panel a) and panel b) are adapted from Figure 15 and 17 in \citet{Limaye2017}, respectively.}
    \label{Day_data_VGCM}
\end{figure}

Beside the high variability observed even on short timescales, a succession of warm and cool layers above 80~km is one of remarkable feature that has been systematically detected. This was also correctly captured by model simulations \citep{Gilli2017, Brecht2012a, Bougher2015} and interpreted as a combination of radiative and dynamical effects: the local maximum at daytime is produced by solar absorption in the non-LTE CO$_2$ IR bands, and then advected to the terminator, while a nighttime warm layer around 90~km is produced by subsidence of day-to-night air at the AS point. Above, the cold layer around 125-130 km is suggested to be caused by CO$_2$ 15~$\mu$m cooling. Nevertheless, especially above 100~km, temperature variation is large and the difference between individual experiments seems to be higher. 
Figures \ref{Night_data_VGCM}-\ref{Day_data_VGCM} are adapted from Figures 16-19 in Limaye17 and show a selection of observed temperature profiles together with IPSL-VGCM simulations on daytime (local time LT= 7h-17h), nighttime (LT= 19h-7h) and at the terminator regions (LT $\approx$ 18h and 6h), averaged in latitude bins. Here we show only equatorial region (0$^\circ$-30$^\circ$) and middle/high (50$^\circ$-70$^\circ$) latitudes, as examples. 
Figures \ref{Night_data_VGCM} and \ref{Terminator_data_VGCM} include retrieved temperature profiles from several instruments on board Venus Express binned in a similar latitude/local time range. Both channels of the Visible and Infrared Imaging Spectrometer (VIRTIS) are used here: VIRTIS-M temperature retrieved using CO$_2$ 4.3~$\mu$m band reported by \citep{Grassi2014, Haus2015} and VIRTIS-H profiles as in \citet{Migliorini2012}. Data from solar and stellar occultation experiment on-board VEx, i.e. SOIR \citet{Piccialli2015} and SPICAV \citet{Mahieux2015}, respectively, from Radio occultation measurements \citep{Tellmann2009} and from O$_2$ nightglow observations at low latitudes \citep{Piccioni2009} are added. For daytime, only temperatures retrieved from non-LTE CO$_2$ emissions measured by VIRTIS/VEx \citep{Gilli2015} are available above 110 km, as shown in Figure \ref{Day_data_VGCM}. Mean temperatures derived from the sunlight refraction in the mesosphere during the 2012 Venus transit \citep{Tanga2012, Pere2016}, and ground-based observations by the JCMT \citep{Clancy2008, Clancy2012}, HHSTM \citep{Rengel2008} and THIS/HIPWAC \citep{Krause2018} are also taken into account. Empirical models such as VIRA \citep{Seiff1985, Keating1985} and VTS3 \citep{Hedin1983} are also plotted as reference. 

As mentioned previously in Sec.~\ref{compar_SOIR-SPICAV}, nightside temperatures (Fig.~\ref{Night_data_VGCM}) predicted by the IPSL-VGCM are still overestimated in the 90-115~km altitude region, by 20 to 30~K, with a peak temperature altitude (around 100~km) that is in agreement with Earth-based observations (e.g. JCMT profiles), but about 5 km altitude higher than the peak of SPICAV temperature retrievals.
Below 90~km, predicted temperatures tend to be at the upper limit of the range of observed values. 
At terminators (Fig.~\ref{Terminator_data_VGCM}), simulated temperature profiles are in good agreement with most of the datasets, except for a cold bias of around 20~K above 130~km, compared to SOIR data. 
The current uncertainty in the collisional rate coefficients used in the non-LTE parameterization (see Section \ref{rec_improv}) is a possible candidate to explain this bias, since K$_{v-t}$ variation has an impact on the temperature above 120 km approximately.

Dayside simulated temperature profiles (Fig.~\ref{Day_data_VGCM}) reproduce the warm and cold layers observed near 110~km and 125~km, respectively, but tend to overestimate their amplitude. Note that the VIRA and VTS3 models, that were under-constrained in the 100-150~km altitude range, do not predict these temperature inversions. Nevertheless, it is difficult to get any conclusion from Figure \ref{Day_data_VGCM}, given the high error bars of daytime temperature measurements obtained with VIRTIS/VEx. 

\begin{figure}[!htbp]
\centering
 \includegraphics[width=0.45\linewidth]{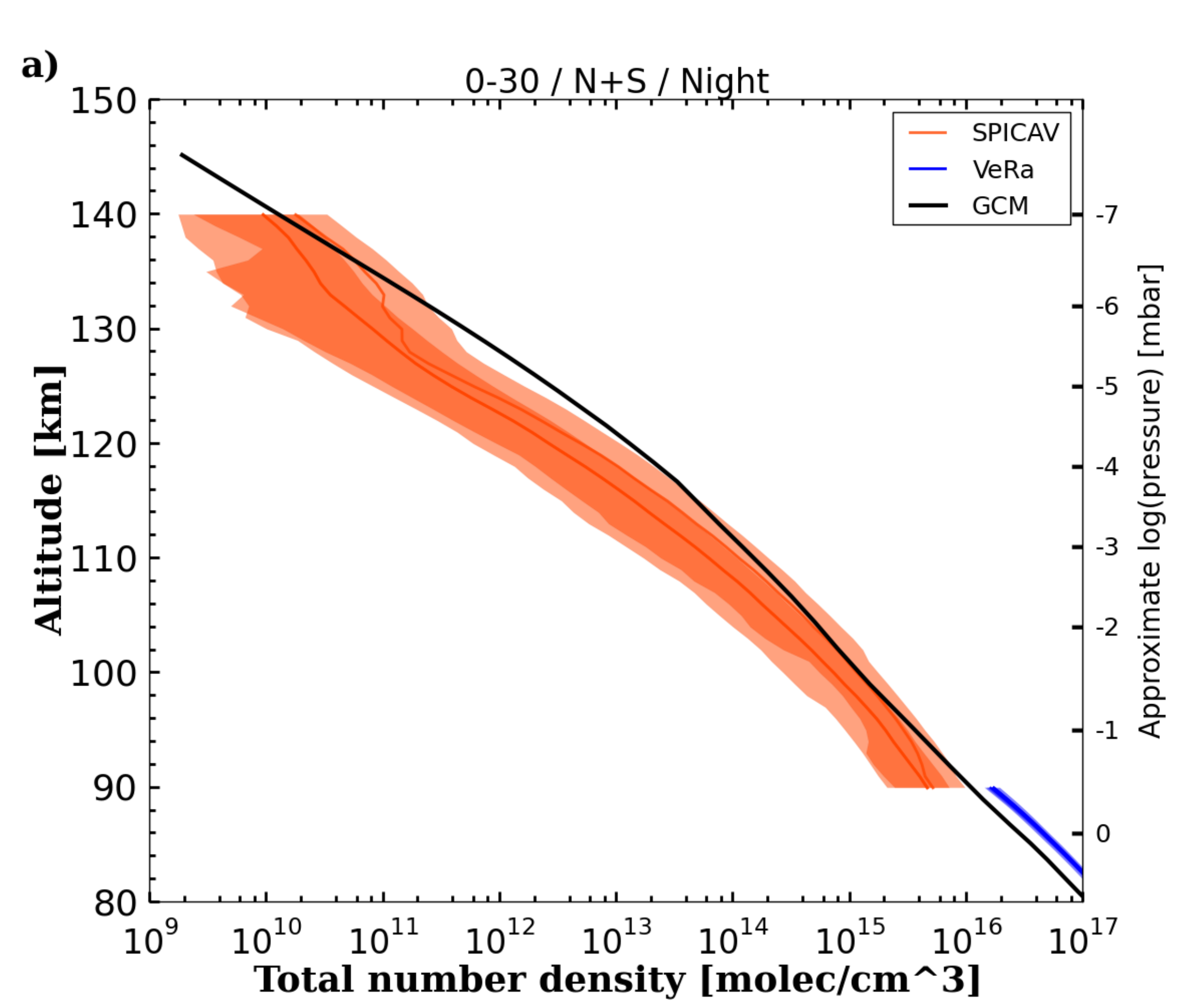}
\includegraphics[width=0.45\linewidth]{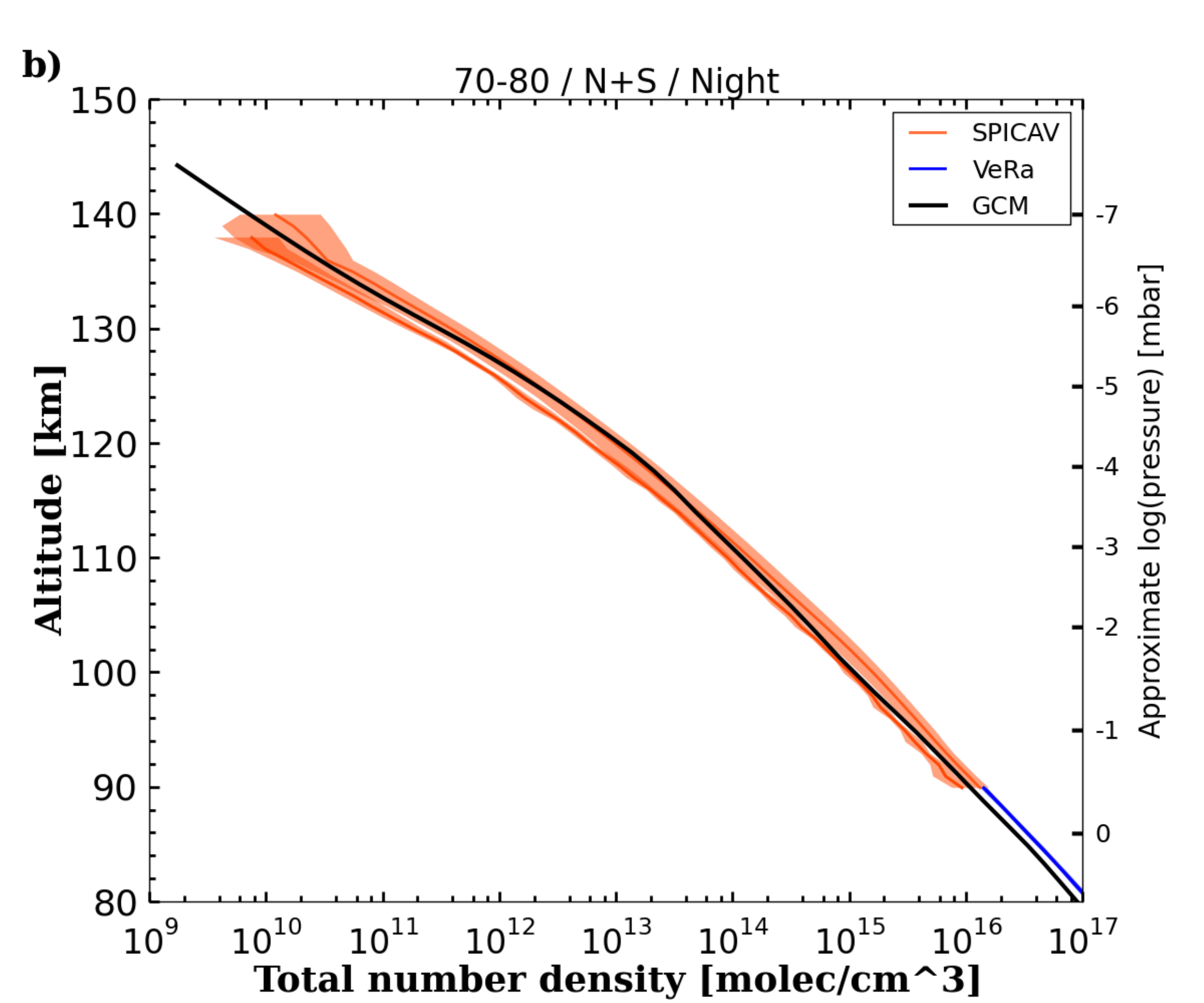}
\includegraphics[width=0.45\linewidth]{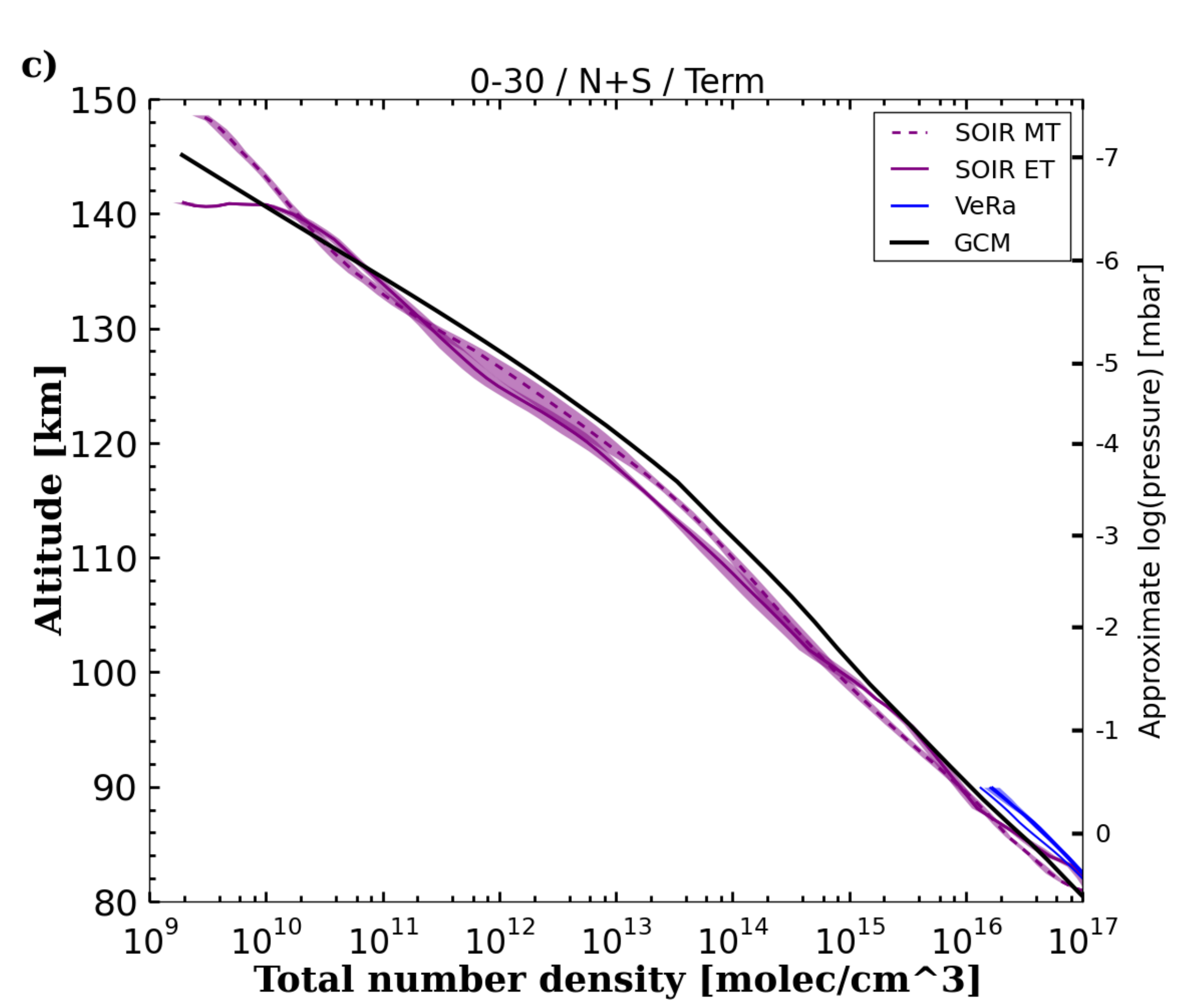}
\includegraphics[width=0.45\linewidth]{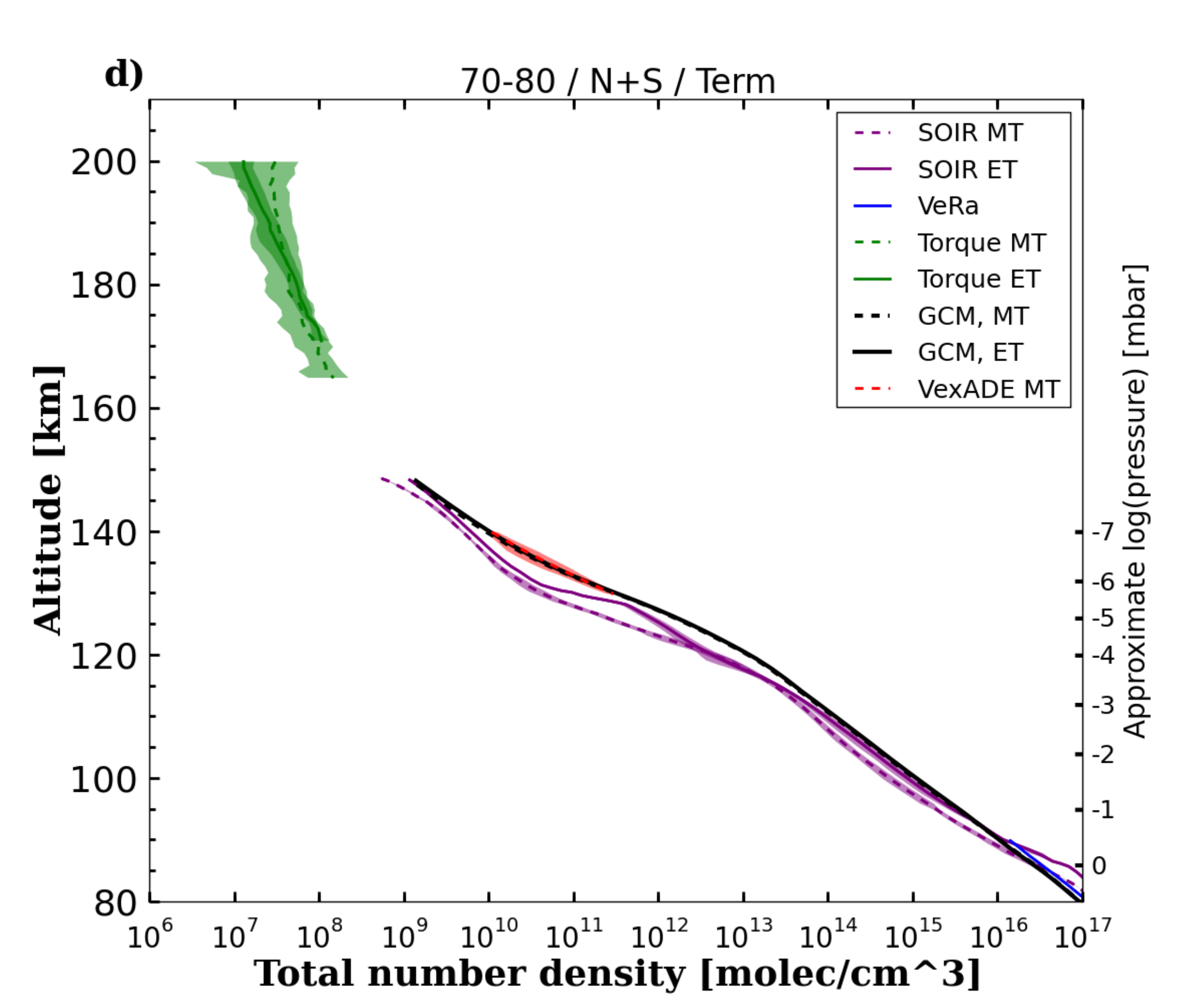}
    \caption{Measured total density profiles (molecule/cm$^3$) retrieved with different Venus Express instruments at nighttime (Panels a and b) and at the terminators (Panels c and d) for two latitude bins 0-30$^\circ$ and 70$^\circ$-80$^\circ$ (North/South averages) as indicated. The colored areas mark the uncertainty of the respective averaged profiles at one standard deviation. The figures are adapted from Figures 20 and 23 in \citet{Limaye2017}. Model predictions averaged at same latitudes and local time as observations are plotted with solid and dashed black lines. Vertical axis is given in altitudes, and approximate pressure (in mbar) is shown as the right hand side vertical axis.}
    \label{CO2_density}
\end{figure}

Retrieved vertical total density profiles from 80~km to 150~km are shown in Figure \ref{CO2_density} as measured by the different Venus Express experiments mentioned above. In addition, density measurements from Venus Express Atmospheric Drag Experiment (VExADE) are included for high latitudes (panel \textit{d} of Figure \ref{CO2_density}): between 130 and 140~km from accelerometer readings during aerobreaking  \citep{Muller-Wodarg2016}, and in the altitude range 165-200~km from spacecraft torque measurements \citep{Persson2015}.
Nighttime densities are in very good agreement with SPICAV and VeRa, with the exception of the equatorial region, where IPSL-VGCM densities are a factor of 2-3 greater than SPICAV observations between 120 and 130~km. Densities retrieved by SOIR at the terminator are also up to a factor 2-3 lower then our simulations in the equatorial region, and up to 1 order of magnitude lower in the altitude between 100-140~km at high latitudes. However, in this region, it is worth noticing that our model predicted values fit very well with the VExADE density profile. 
The discrepancy with SOIR may be linked to the assumption of spherical homogeneous layers made by the authors and on the a-priori CO$_2$ mixing ratio. SOIR and SPICAV measured directly the CO$_2$ number density from its absorption structure, thus have to assume a CO$_2$ volume mixing ratio, taken from VTS3 semi-empirical model \citep{Hedin1983} above 100~km. VTS3 is mainly based on PV measurements above 140~km and on model extrapolations assuming hydrostatic equilibrium between 100-140~km. This model revealed not to be adapted to represent the atomic oxygen densities derived from O$_2$ night-glow observations \citep{Soret2012}, the measured value being 3 times higher than predicted by VTS3. Moreover, the observed high variability of atmospheric quantities by VEx instruments was not reflected in previous empirical models (e.g. VIRA and VTS3) \citep{Limaye2017}. 
Therefore, the uncertainties of SOIR profiles may reflect uncertainties in the VTS3 densities and have to be taken into account when comparing the observations with models.

\subsection{CO densities}\label{COdens}

%%%&&&&&&&&&&&&&&&&&&&&&&&&&&&&&&&&&&&&&&
%%% Figures  8,9, 10 ,11. DENSITY CO, and O compared with VEX
%%%%%%%%%%%%%%%%%%%%%%%%%%%%%%%%%%%%%%%%%%%%%

\begin{figure}[!htbp]
    \centering
    \includegraphics[width=1.\linewidth]{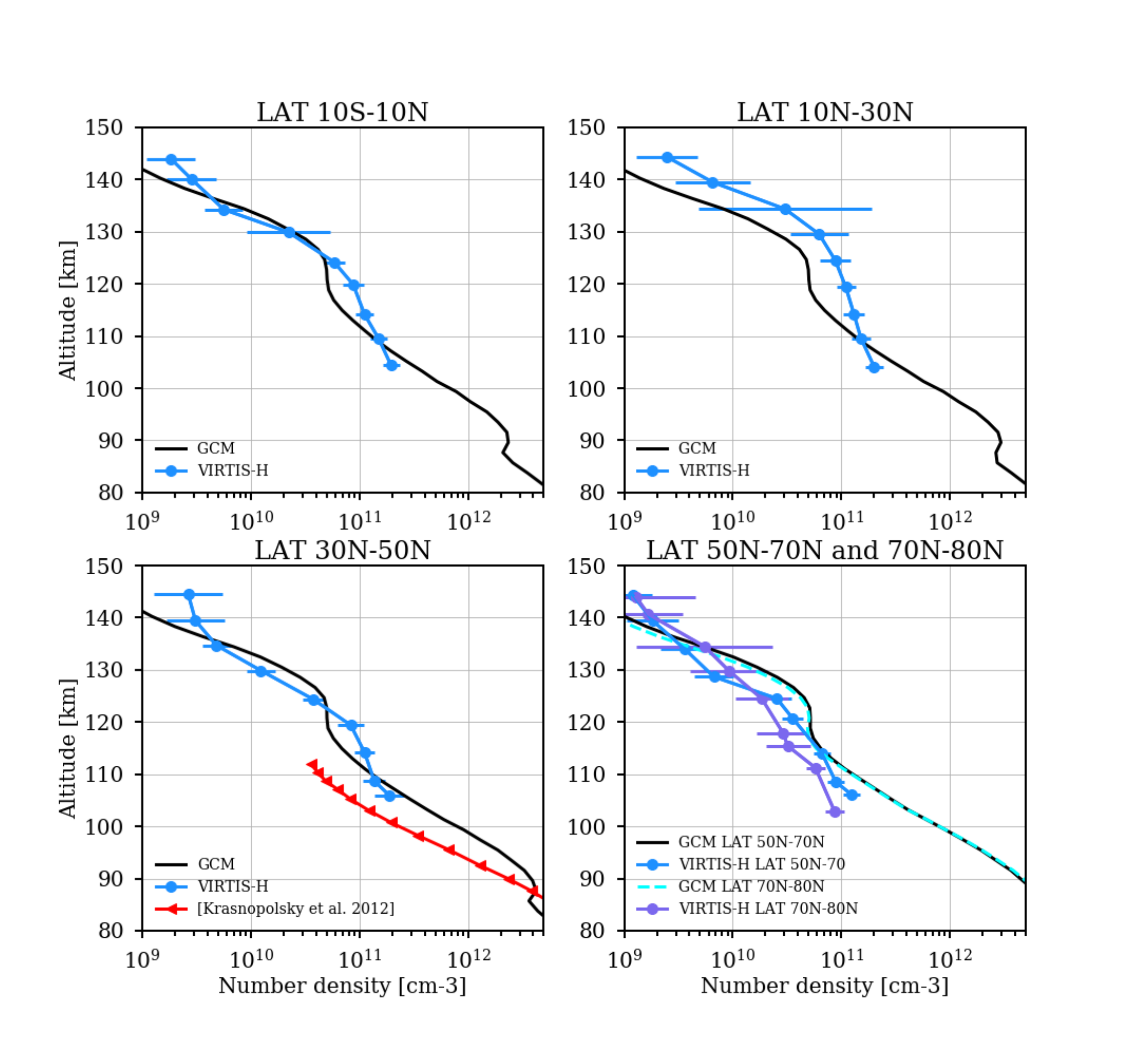}
    \caption{CO density retrieved by VIRTIS/VEx during daytime (LT 10h-14h) after \cite{Gilli2015} together with IPSL-VGCM averaged profiles binned at similar latitude and local time, as indicated in the panels. In addition, semi-empirical CO profile from \citet{Krasnopolsky2012} representative of mean conditions is shown for comparison with mid-latitude (30$^\circ$N-50$^\circ$N) daytime averaged profiles.}
    \label{CO_VIRTIS_day}
\end{figure}

\begin{figure}[!htbp]
    \centering
    \includegraphics[width=1.\linewidth]{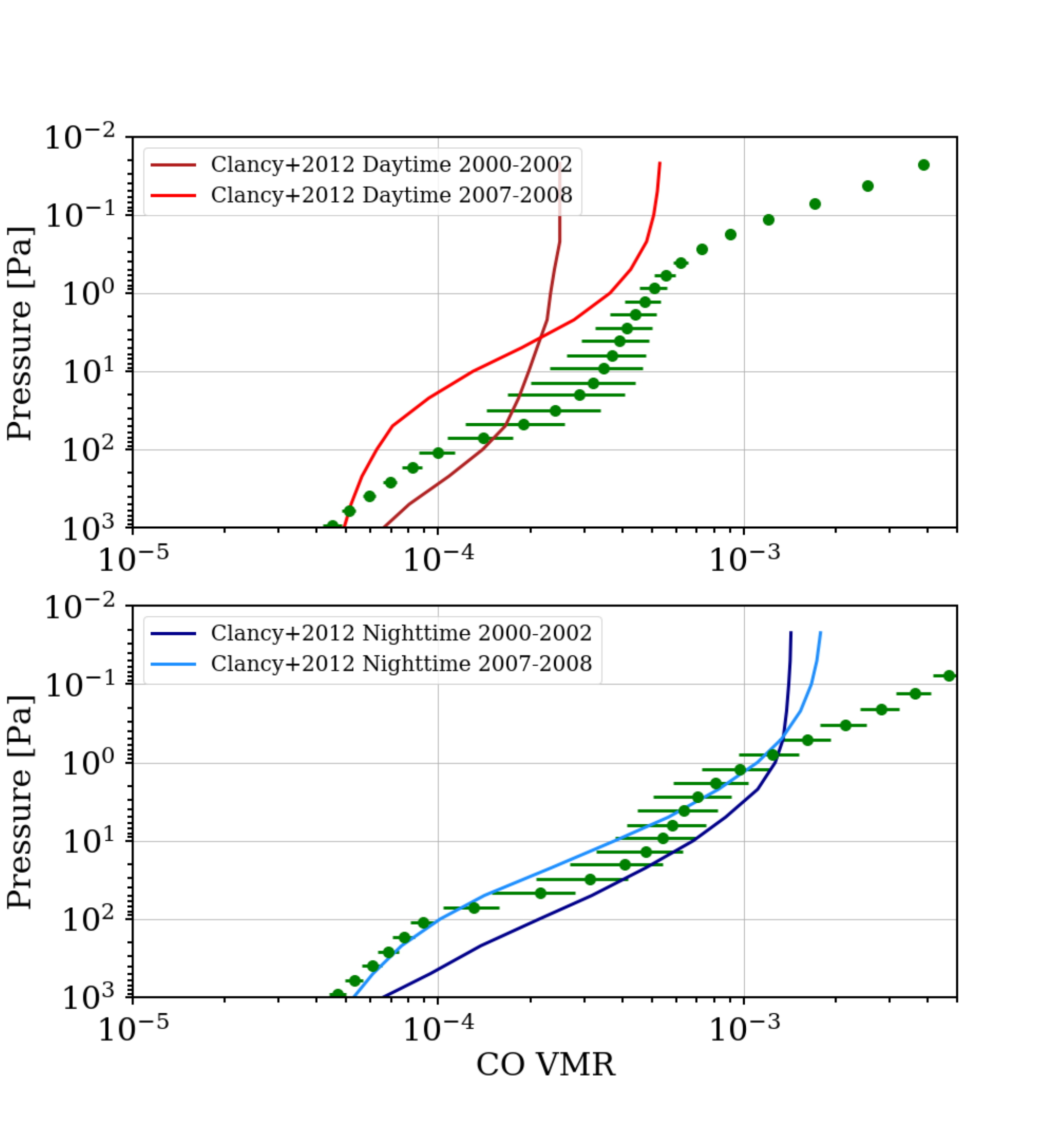}
    \caption{IPSL-VGCM CO volume mixing ratio profiles (green dots) as function of pressure, averaged in latitude bin 60$^\circ$N-60$^\circ$S, 10h-14h LT for daytime and 22h-2h for nighttime, with variability indicated as standard deviation of the average. CO profiles retrieved from sub-mm CO absorption line observations as in \cite{Clancy2012} for daytime (top panel) and nighttime (bottom panel) are also plotted. 2000-2002 disk average daytime/nighttime measurements are in dark red/blue solid lines, while 2007-2008 disk average daytime/nighttime measurements are in red/blue solid line. }
    \label{CO_Clancy}
\end{figure}

\begin{figure}[!htbp]
    \centering
    \includegraphics[width=1.\linewidth]{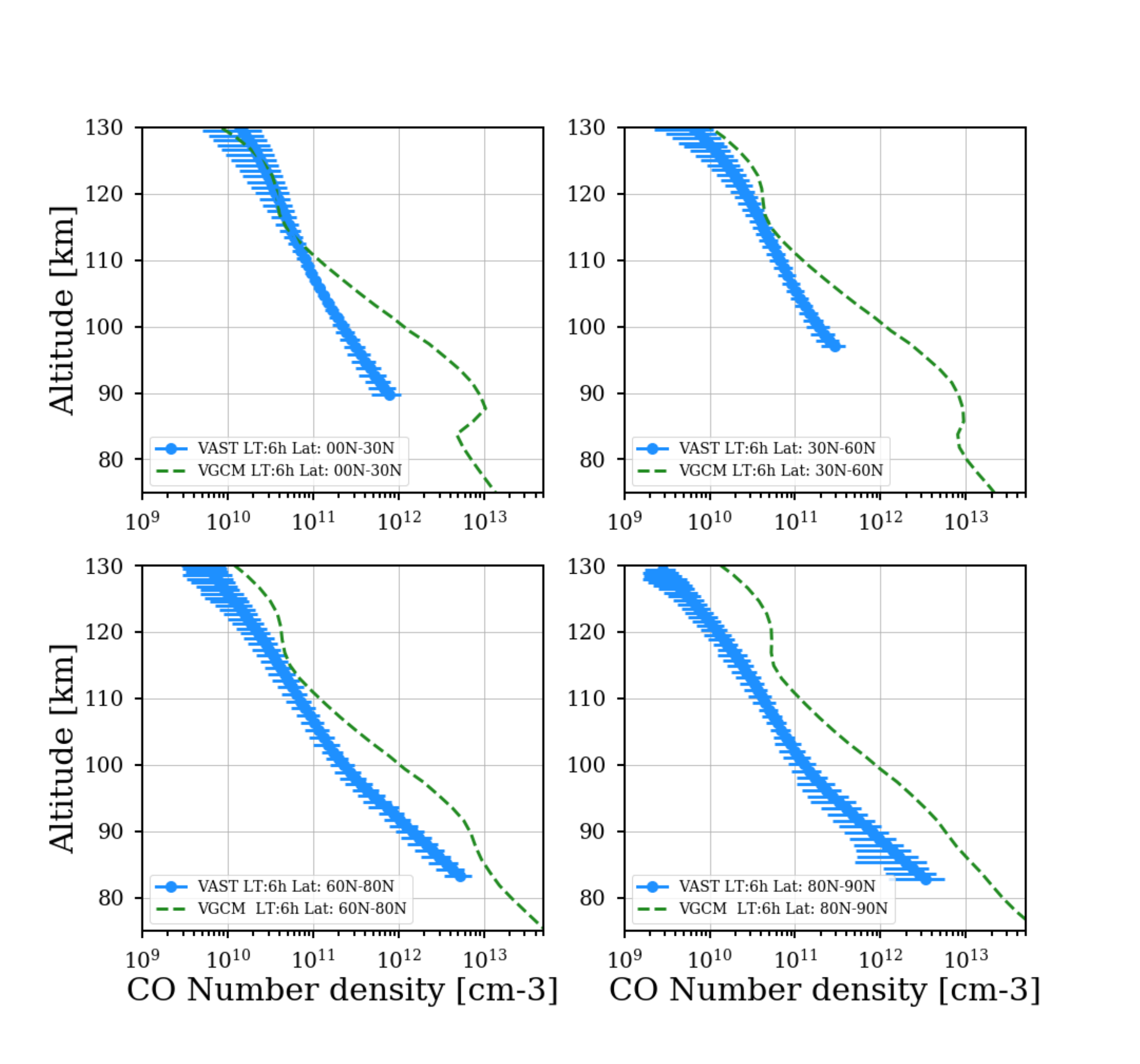}
    \caption{IPSL-VGCM simulated CO density profiles as function of altitude (green dashed lines) at the morning terminator (MT) (LT 6h) averaged at four selected latitude bins as indicated in the panels: 0$^\circ$-30$^\circ$N, 30$^\circ$N-40$^\circ$N, 60$^\circ$N-80$^\circ$N, 80$^\circ$N-90$^\circ$N.
     CO density profiles and relative error bars retrieved by SOIR/VEx at the MT in the same latitude bins are plotted in blue. VAST stands for: ``Venus Atmosphere from SOIR data at the Terminator", after \cite{Vandaele2016}.}
    \label{CO_SOIR_MT}
\end{figure}

\begin{figure}[!htbp]
    \centering
    \includegraphics[width=1.\linewidth]{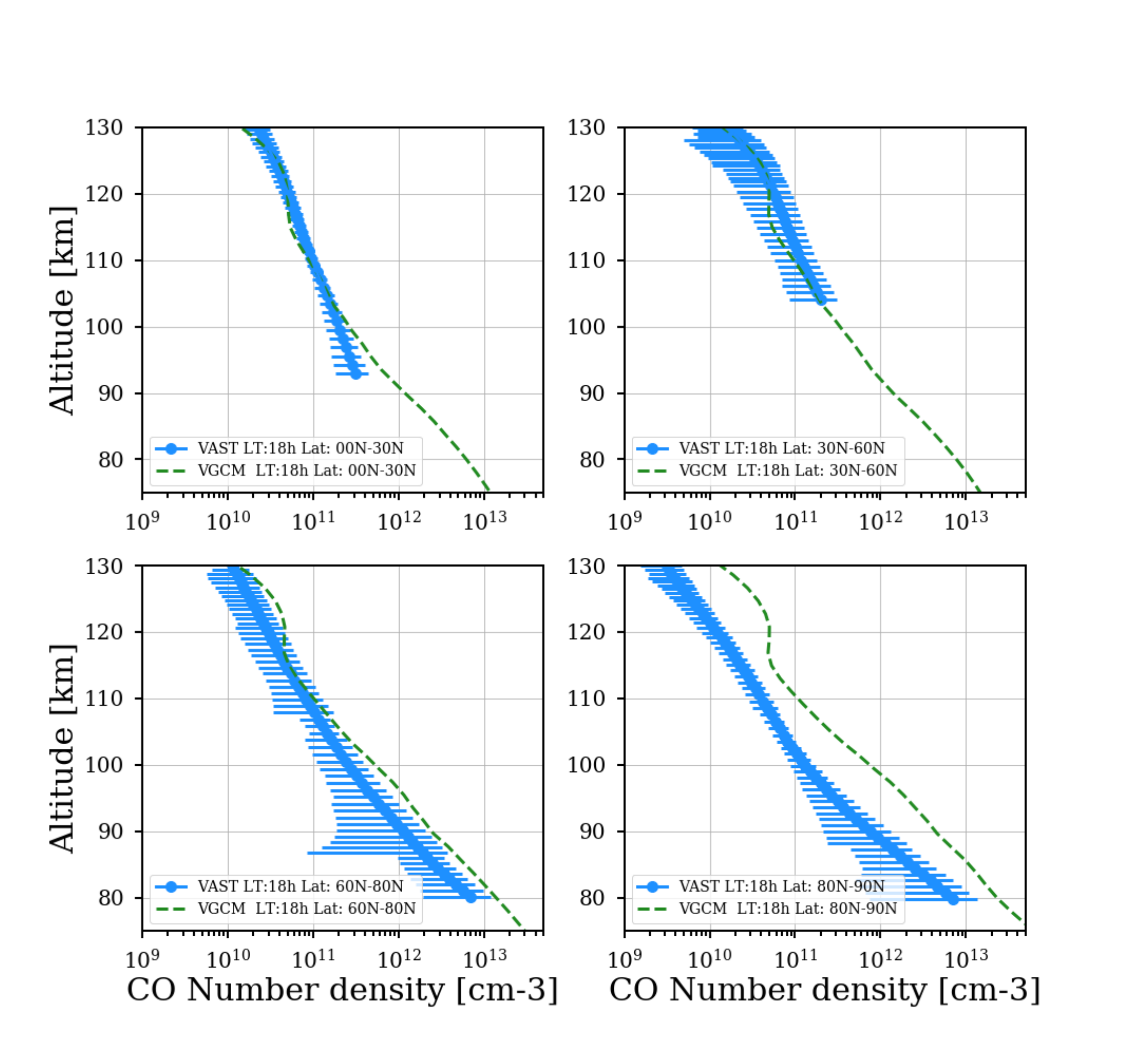}
    \caption{Same as Figure \ref{CO_SOIR_MT} but for evening terminator (ET), LT = 18h}
    \label{CO_SOIR_ET}
\end{figure}

Figures \ref{CO_VIRTIS_day} to \ref{CO_SOIR_ET} show density of CO, compared with a selection of VEx experiments and ground-based measurements at different local time and latitudes.
For daytime, retrieved CO density measurements are very scarce. We used here VIRTIS/VEx measurement in the altitude range 100-150~km as in \cite{Gilli2015} retrieved from non-LTE dayside IR CO emission at 4.7~$\mu$m. Below 112~km and for latitudes between 30$^\circ$N-50$^\circ$N, we also included the CO density profile from the semi-empirical model by \citet{Krasnopolsky2012} which is representative of averaged values under intermediate solar conditions. We found that our results are very well comparable with observations in order of magnitude. However, daytime retrieved CO density between 110 and 125 km are about a factor 2 larger than our model profiles for most latitudes, except for high latitude bins (50$^\circ$N-70$^\circ$N) where instead observed CO densities are smaller than simulations, up to a factor 3. 
Those discrepancies could be related to the biases in the predicted temperature structure that we found above 90~km, the model being systematically warmer than the data between 90-110~km (approximately 10$^{2}$—10$^{-1}$Pa), and colder between 120-150~km. 
%\textbf{Adding error bars: min/max values of the model will add to define this comparison ?}

Regarding the terminators (Figs.~\ref{CO_SOIR_MT}-\ref{CO_SOIR_ET}), our predicted CO densities are in overall agreement with SOIR dataset \citep{Vandaele2016} in particular at evening terminator, but with larger differences with observed values below about 110~km on the morning side. The discrepancies are up to one order of magnitude between 90 and 100~km, and factor 2-3 elsewhere. Larger discrepancies are seen in the polar region (80$^\circ$N-90$^\circ$N), where our model is up to one order of magnitude larger than data for both MT and ET.

 Daytime and nighttime disk averaged CO measurements were also obtained in several ground-based campaigns \citep{deBergh1988, ClancyMuhleman1991, Lellouch1989_CO,Clancy2008, Clancy2012} suggesting not only large variability over daily to weekly timescales, but also a distinct dayside versus nightside circulation in terms of zonal wind in particular \citep{Clancy2012}.
Figure \ref{CO_Clancy} shows the comparison of model outputs and CO volume mixing ratio measurements during ground-based campaigns described in \cite{Clancy2012}. They represent 2001-2009 inferior conjunction sub-mm CO line observations, covering most of the disk (e.g. 60$^\circ$S-60$^\circ$N latitudes). Daytime model outputs are averaged for LT 10h-14h, and nighttime for LT 2h-22h, in the observed latitudinal bin. Our results are in general good agreement with observations for pressure ranges 10$^3$-10 Pa, but above that pressure the trends are diverging (i.e. increasing volume mixing ratio (vmr) predicted by the model versus almost constant CO values suggested by observations). Nighttime measurements of the 2007-2008 campaign are within model dispersion bars (calculated as the standard deviation in the selected latitude/local time bin), and show a better agreement than 2000-2002 campaigns.  Part of the difference between the two campaigns is likely related to the much reduced beam coverage for those early observations (Clancy, private communication).

Given the high variability of CO measurements over short time scales, it is difficult to find a general consensus among observations, and a clear defined trend. Our simulations indicate that the predicted abundances are affected by both the asymmetry of the SS-AS zonal flow at the terminator produced by non-orographic GW propagation in the lower thermosphere and the variability of zonal winds induced by Kelvin waves generated in the cloud deck, and propagating up to the mesosphere, as discussed in Navarro21. A possible interpretation of CO distribution is also provided in Section \ref{Discussion}. 

\subsection{O densities}

Only nighttime oxygen density measurements are available in the UMLT of Venus and they are derived from O$_2$ nightglow in the altitude range 85-110 km observed with VIRTIS/VEX, as described in \cite{Soret2012}. Figure \ref{Odens_Soret_Krasno} shows IPSL-VGCM results together with the observed profile near the equator (24$^\circ$S) and nighttime O density extracted from a semi-empirical model by \citet{Krasnopolsky2012}.
In very good agreement with \cite{Soret2012}, modeled values peak at 2$\times$10$^{11}$ cm$^{-3}$ around 110 km. \cite{Brecht2011} also found a similar value at midnight in their thermospheric GCM simulations at lower altitudes (3.4$\times$10$^{11}$ cm$^{-3}$ at 104 km). 
Below 95~km, O densities predicted by our model are decreasing much faster with altitude than observed in \cite{Soret2014}, but this is still consistent with \citet{Krasnopolsky2012}. A more complete description of the latitudinal and diurnal distribution of O predicted by the IPSL-VGCM is presented in the next section. 

%%%%%%%%%%%%%%%%%%%%%%%%%%%%%%%%%%
%%%. Sect. 4
%%. 
\section{Diurnal and latitudinal variations of CO and O above 80 km by the IPSL-VGCM}
\label{Diurnal_Lat_variation}
The latitudinal and diurnal variations of CO and O predicted by the IPSL-VGCM are analysed here in greater detail.
Atomic oxygen chemical lifetime is shorter than carbon monoxide. Therefore the O mixing ratio decreases more rapidly with increasing pressure than CO, and its latitudinal variations are not significant in the GCM above $\sim$10 Pa (95 km approximately). This is clearly shown in Figures \ref{lat_var_day}-\ref{lat_var_night}: CO and O \textit{vmr} profiles by IPSL-VGCM as function of pressure are plotted for several latitude bins, at daytime (LT: 10h-14h) and nighttime (LT: 22h-2h). 
%%%% CO variation
\\In the lower mesosphere CO variations are primarily dominated by the global meridional dynamics, with CO \textit{vmr} increasing from equator-to-pole \citep{Tsang2008, Marcq2008}. At high latitudes, downward fluxes from the Hadley-type cell circulation bring air enriched in CO, whereas the ascending branch of these cells has the opposite effect near the equator \citep{Marcq2008}. This equator-to-pole variation is predicted by our model below 1 Pa ($\sim$ 97-100~km), while oxygen \textit{vmr} variation is negligible.
In the upper mesosphere (above 1 Pa), CO as other light species is expected to pile up at the converging stagnation point of the wind field (i.e. where the horizontal velocity converges to zero). 
%This point is at the anti-solar point for dominating SS-AS flow and displaced toward the morning terminator when a retrograde zonal flow is added. 
%The position of the maximum and its magnitude depend on the relative values of equatorial velocity and a maximum cross-terminator velocity.
This point can be displaced from the anti-solar point when the circulation is not simply the SS-AS flow.
Our simulations show that the CO nighttime bulge is shifted toward the morning between 85 and 100~km approximately, notably around the equator (see Figure~\ref{CO_O_LT_var}), suggesting that CO is preferentially transported westwards in the transition region. Interestingly, \citet{Lellouch2008} observations showed a moderate (factor 2) CO enhancement near the morning  terminator, although there are no systematic observations supporting this enhancement of CO on the morning side.

\begin{figure}[!htbp]
    \centering
    \includegraphics[width=1.\linewidth]{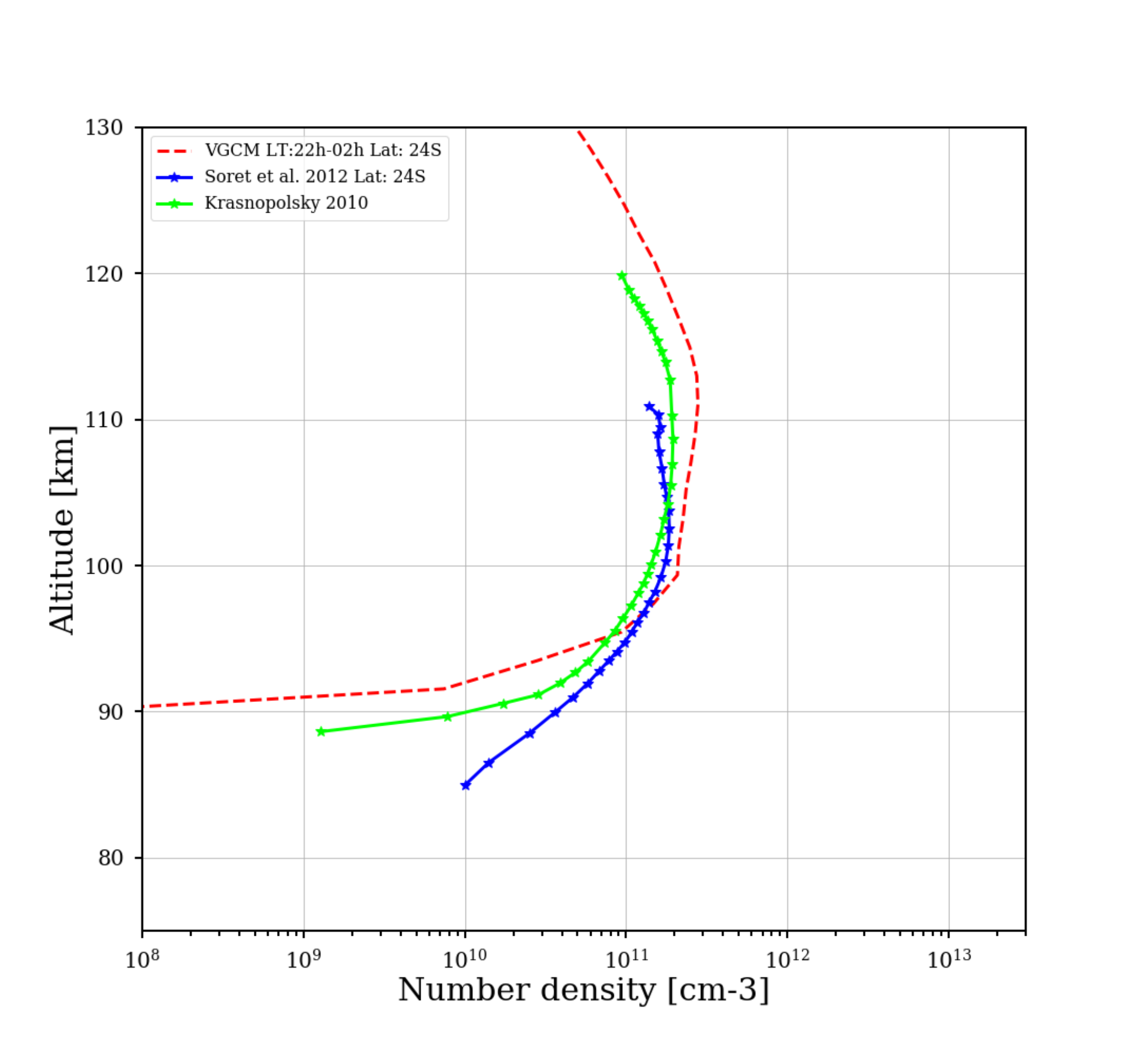}
    \caption{Nigthtime O density retrieved by VIRTIS/VEx from O$_2$ nightglow observations at the equator (Lat: 24$^\circ$S) as in \cite{Soret2012} (blue line) and from \citet{Krasnopolsky2012} semi-empirical model (green line). IPSL-VGCM simulated nighttime profile (LT: 22h-2h) for the same latitude is shown with red dashed line.}
    \label{Odens_Soret_Krasno}
\end{figure}

\begin{figure}[!htbp]
    \includegraphics[width=1\linewidth]{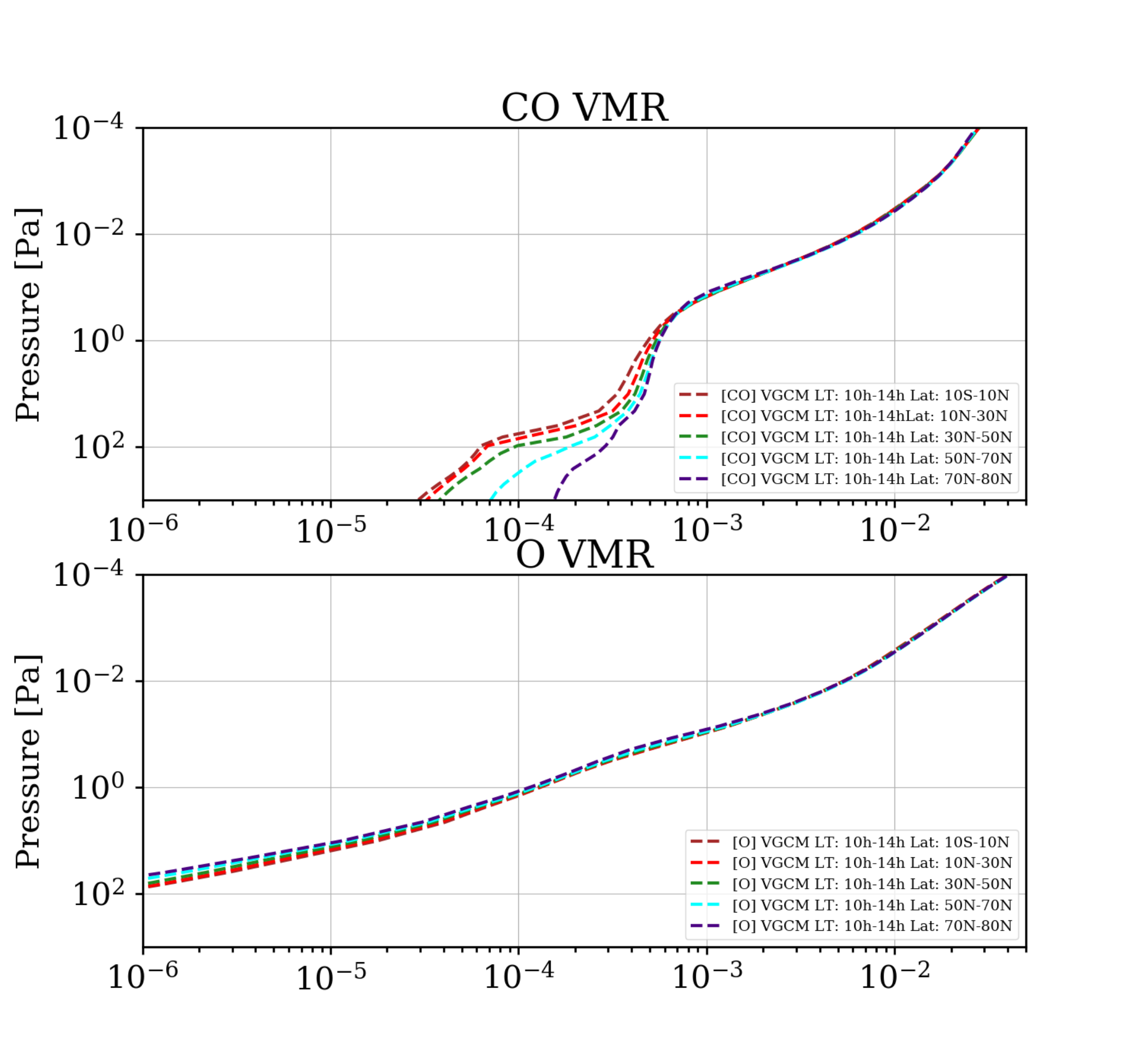}
    \caption{Latitudinal variations of CO and O vmr during daytime (10h-14h) predicted by the IPSL-VGCM. The latitude bins are indicated in each panel. The pressure levels correspond to an altitude range between 75 and 140 km, approximately. CO profiles show a clear transition of regime at around 1 Pa (100-110 km)}
    \label{lat_var_day}
\end{figure}

\begin{figure}[!htbp]
        \includegraphics[width=1\linewidth]{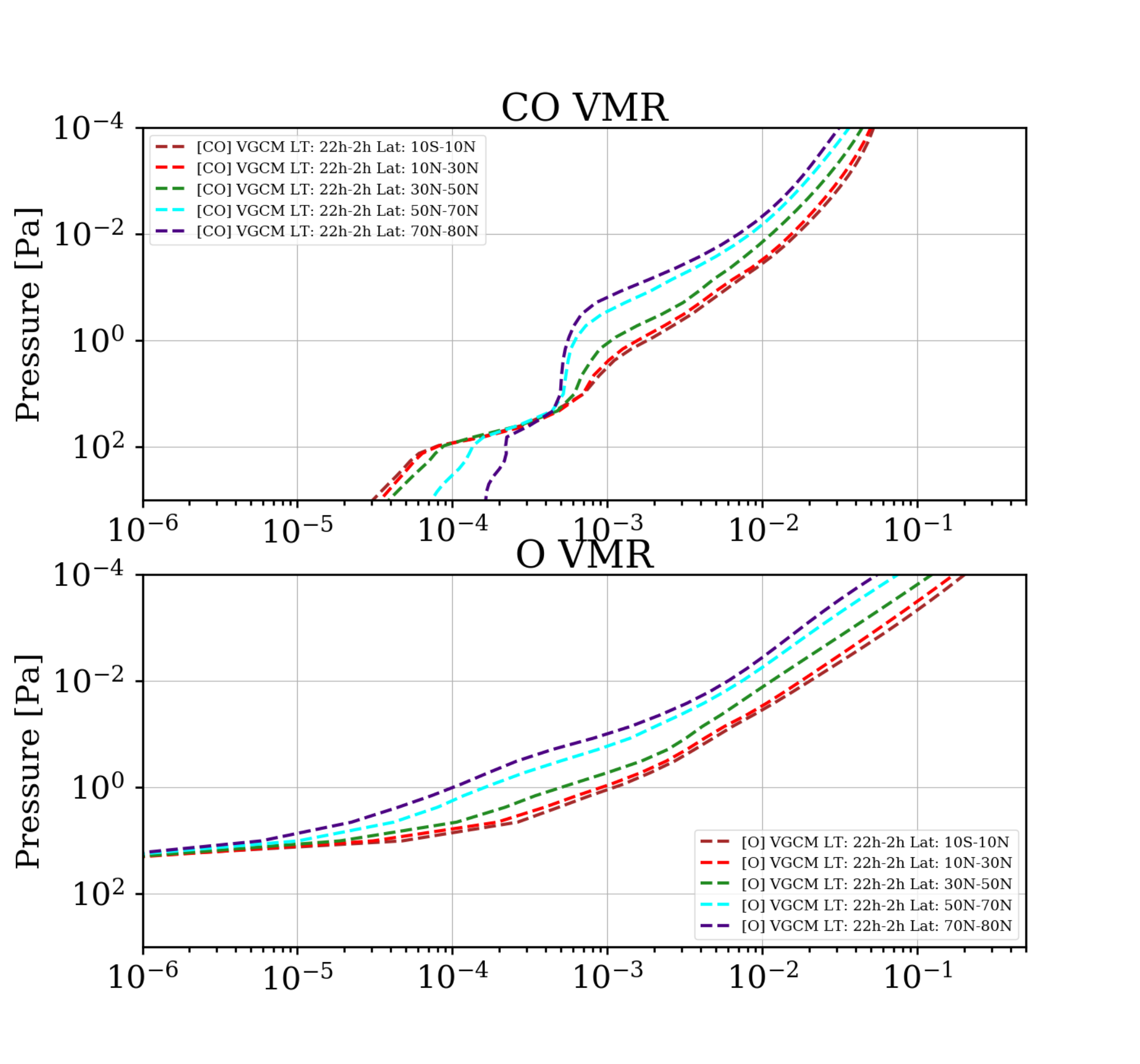}
    \caption{Latitudinal variations of CO and O vmr during nighttime predicted by the IPSL-VGCM. The latitude bins are indicated in each panel.}
    \label{lat_var_night}
\end{figure}

\begin{figure}[!htbp]
    \centering
    \includegraphics[width=0.8\linewidth]{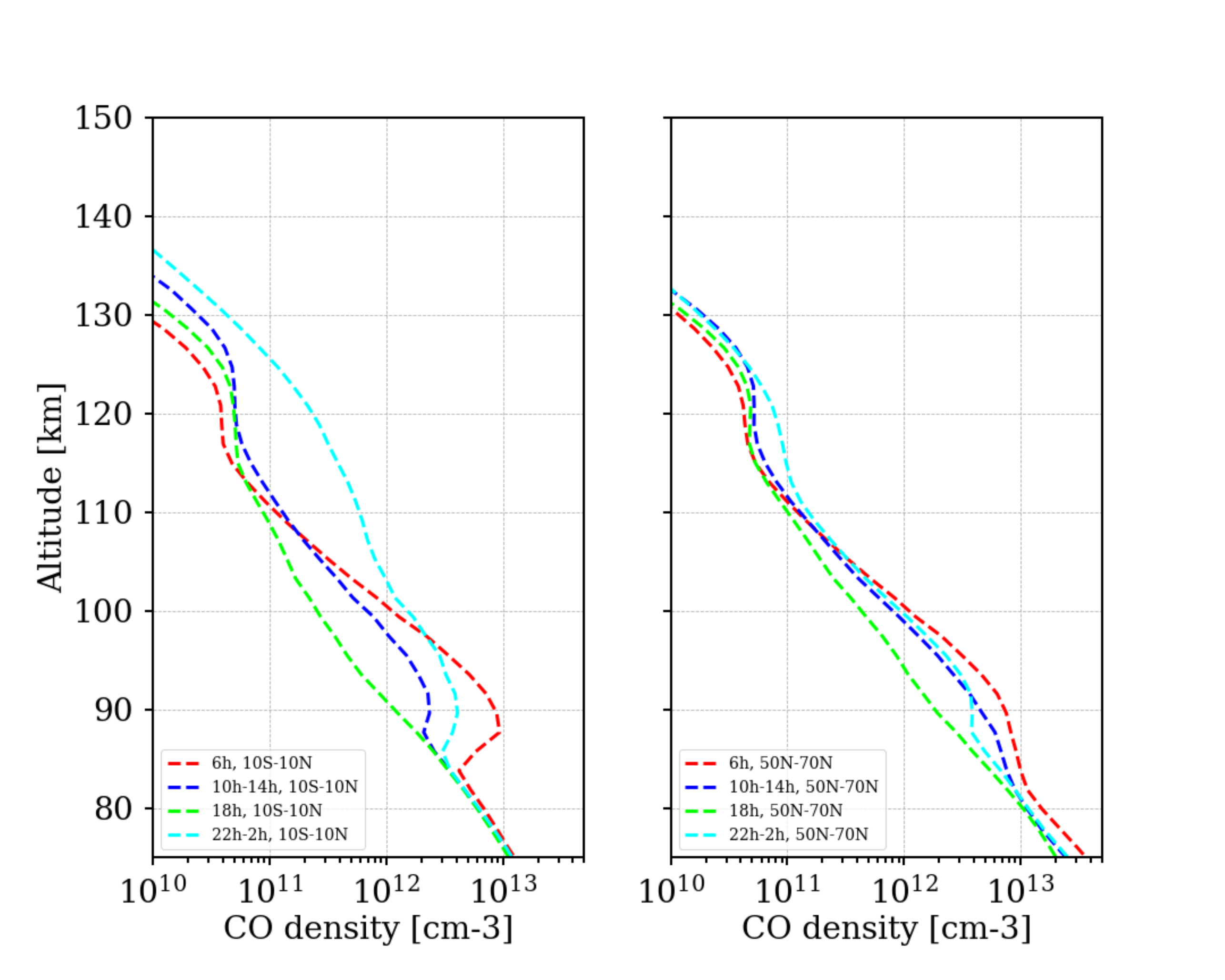}
    \includegraphics[width=0.8\linewidth]{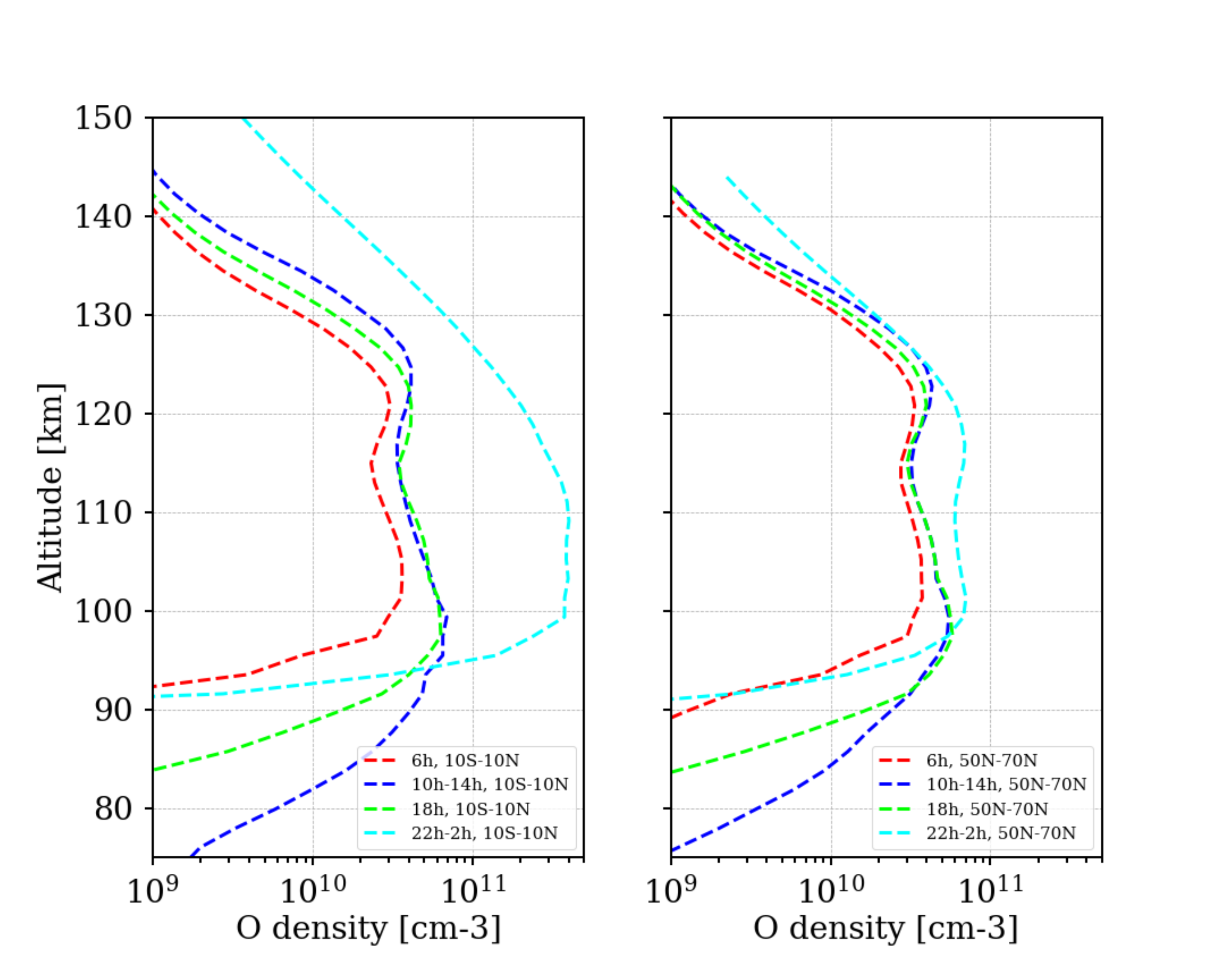}

    \caption{Local time variations of IPSL-VGCM CO (top panels) and O (bottom panels) density profiles at equatorial regions 10S-10N (left panels) and middle-high latitudes 50N-70N (right panels). Daytime and nighttime profiles are averaged between LT=10h-14h and LT=22h-2h, respectively. Evening (18h) and morning (6h) terminators are also plotted.}
    \label{CO_O_LT_var}
\end{figure}

%%%%% O Variation
O chemical lifetime is shorter than the characteristic times of the atmospheric dynamics below 1~Pa approximately, where changes in \textit{vmr} are driven by chemistry. The opposite is expected above 100~km, where the dynamical effects dominate \citep{Brecht2011}. The atomic oxygen density peaks at midnight at the equator between 100 and 110~km, and it is about 1 order of magnitude larger than density values at the same altitude, at noon. The daytime peak is also located below 100~km. This enhancement from day to night side was also predicted in the Venus GCM by \citet{Brecht2011}(e.g. a factor of 6 larger), and it is the result of efficient transport of atomic oxygen atoms from their day sources to their nightside chemical loss at and below about 110 km.  
In addition, our model predicts a second peak around 120~km (Figure \ref{CO_O_LT_var}), clearly defined for local time 10h-14h and at terminator, but weaker at midnight. Unfortunately, no observational data can be used to validate those trends for daytime and terminator. Only O density retrieved indirectly from O$_2$ nightlow are available so far, and they are in very good agreement with our prediction (see Figure \ref{Odens_Soret_Krasno}). The origin of this second peak is beyond the scope of this paper and deserves a dedicated theoretical study, supported by systematic measurements of O densities. %Interestingly, the presence of a double peak of O$_2$ nightglow profiles was detected from VIRTIS/VEx limb observations between 80 and 120~km in \cite{Piccioni2009}, and interpreted as propagation of gravity waves with vertical wavelength of 10-14 km, and horizontal wavelength of 100-1000~km \citep{Altieri2014}.

%%%%%%%%%%%%%%%%%%%%%%%%%%%%%%%%%%
%%%. Sect. 5
%%.
\section{Discussion: dynamical implications}
\label{Discussion}
%\textit{Impact of Kelvin-like wave:} 
\subsection{Impact of the Kelvin wave} \label{KWimpact}
Several studies \citep{Hueso2008,Clancy2012,Gerard2014,Lellouch2008, Moullet2012} indicated that both retrograde zonal and SS-AS flows affect the global distribution of CO, O and other light species at mesospheric altitudes (70-120 km). Those chemical tracers are transported by winds and are expected to pile up at the converging stagnation point of the wind. This point is usually at midnight (LT=0h) for pure SS-AS flow but it can be displaced toward the morning terminator if westward retrograde zonal flow is added. The position of the maximum and its magnitude depend on the relative values of equatorial velocity and a maximum cross-terminator velocity.
The majority of Doppler winds retrieved at mesospheric and lower thermospheric altitudes \citep{Clancy2008,Clancy2012,Rengel2008, Widemann2007, Gurwell1995} suggested local time variation of the wind velocity and the presence of a substantial retrograde zonal flow at altitudes above 90 km. \citet{Lellouch2008} observations showed a moderate (factor of 2) CO enhancement near the morning  terminator, and \citet{Gurwell1995} showed a CO maximum centered at 3h30 at 95 km, and at 2h at 100 km, approaching 2$\times$10$^{-3}$ volume mixing ratio.

\begin{figure}[!htbp]
%    \centering
   \includegraphics[width=0.5\linewidth]{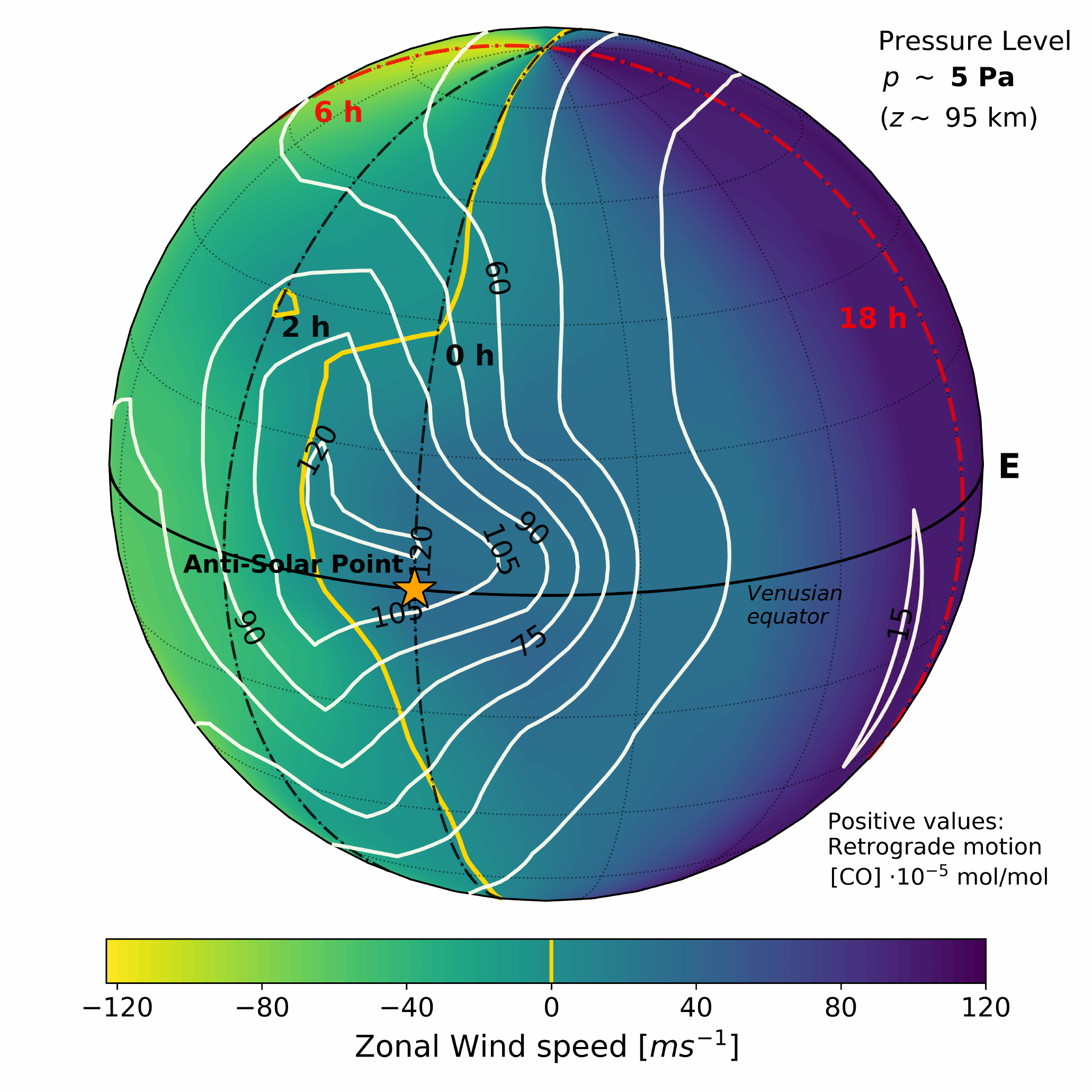}
    \includegraphics[width=0.5\linewidth]{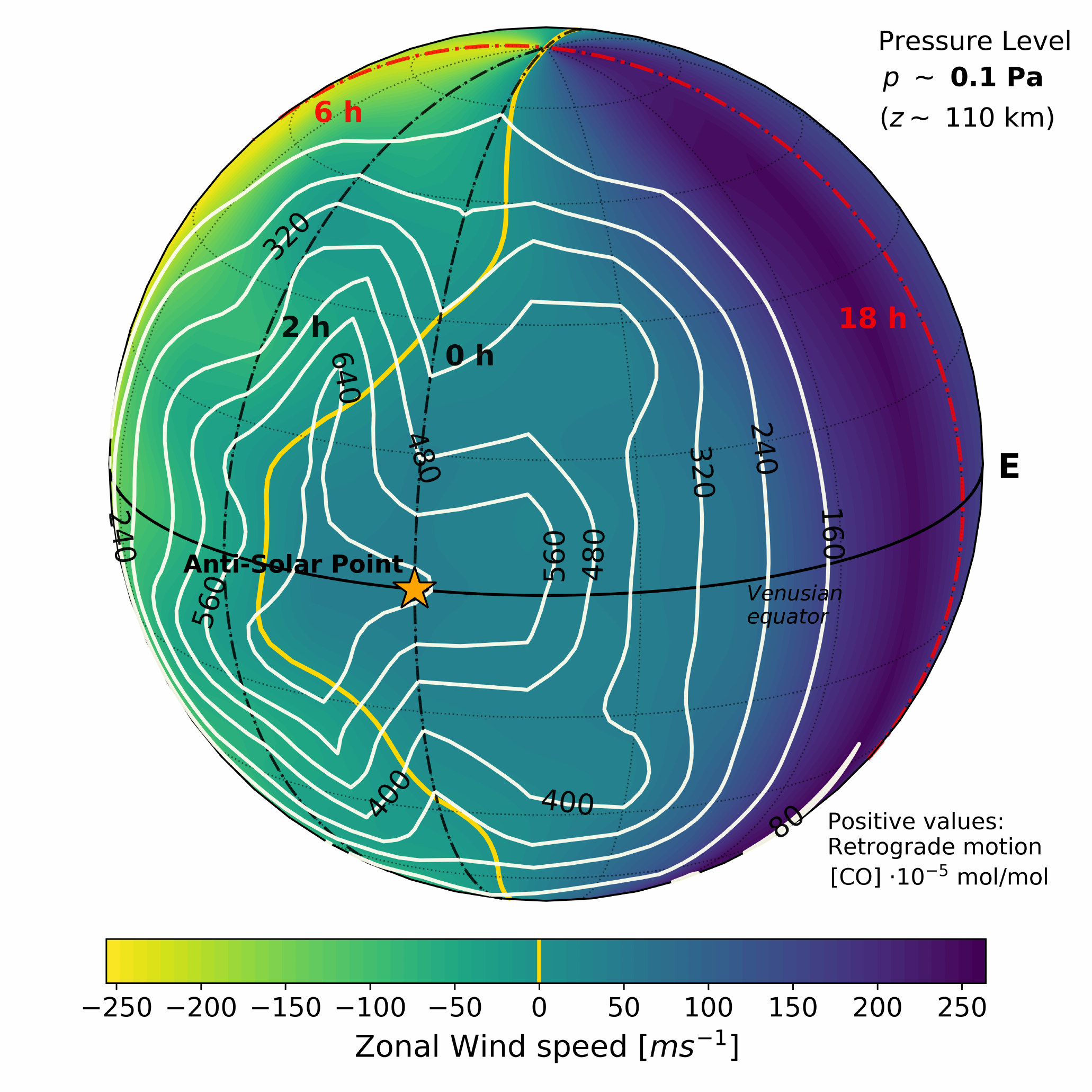}
    \caption{3D maps of Venus nightside at 5 Pa (left) and 0.1 Pa (right) averaged for one Venus solar day, the anti-solar point is indicated with a star. Contour white lines: CO vmr and colors are zonal wind fields. Yellow lines indicate where the horizontal wind converges to zero values. The location of the terminator is represented by red dashed lines.}
    \label{3D_map_CO_u}
\end{figure}

\begin{figure}[!hbtp]
%    \centering
   \includegraphics[width=1.1\linewidth]{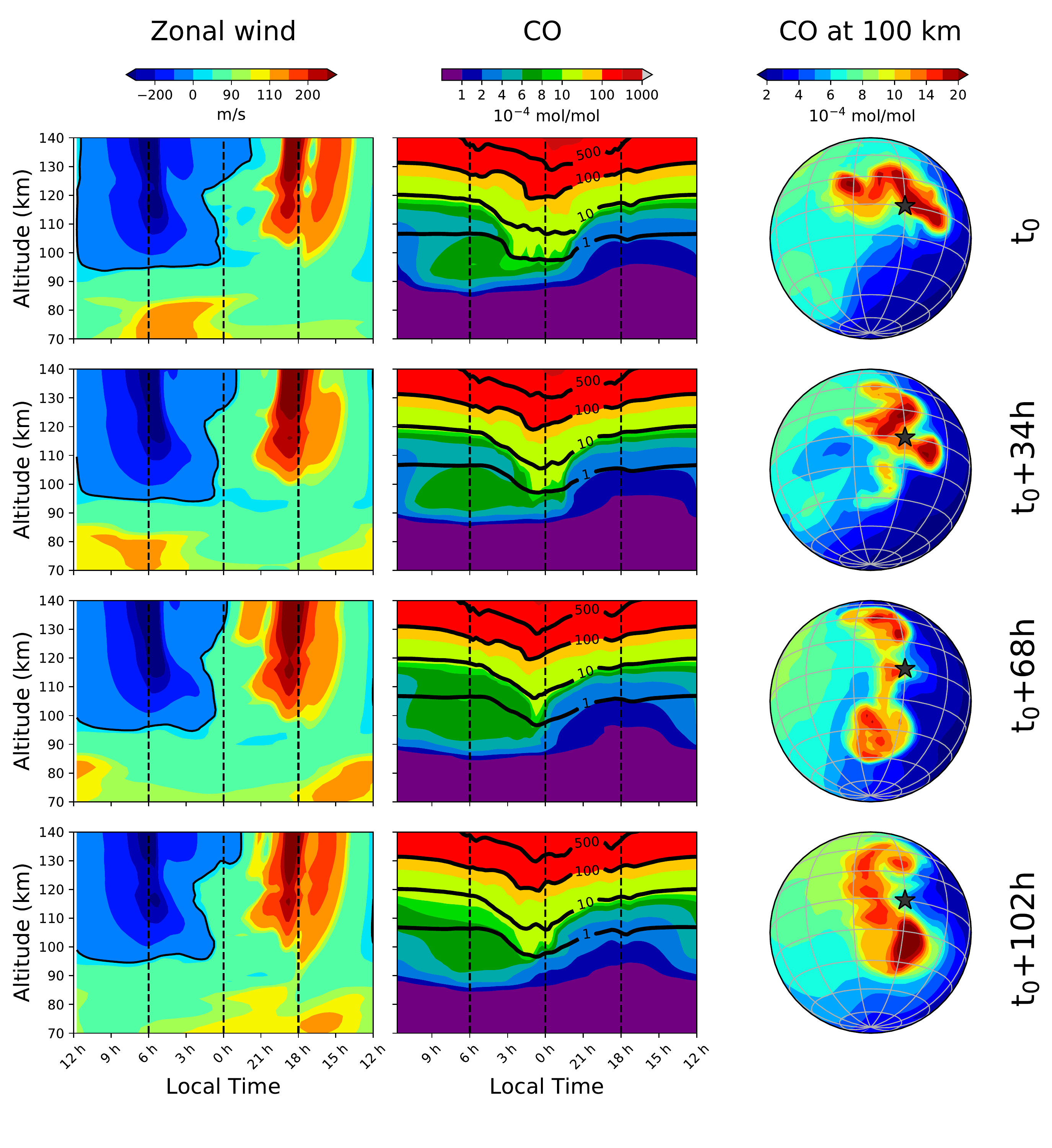}
    \caption{Four consecutive snapshots, from left to right, of: (1) 2D local time vs. altitude map of zonal wind averaged for latitudes 20$^\circ$S-20$^\circ$N, (2) 2D local time vs. altitude map of CO vmr [mol/mol] averaged for latitudes 20$^\circ$S-20$^\circ$N (3) 3D map of CO vmr [mol/mol] at 100 km, the AS point is indicated by a black star.}
    \label{2D_maps_CO_TN}
\end{figure}

This retrograde imbalance of the flow towards the morning side is also predicted by our model. CO nighttime bulge is not located at the AS-point but shifted toward the morning (see Figure~\ref{3D_map_CO_u} and Figure~\ref{2D_maps_CO_TN}) and in very good agreement with \citet{Gurwell1995}. The companion paper Navarro21 describes how the nightside mesospheric circulation is perturbed by a Kelvin wave with a period of approximately 5 Earth days
In our simulations the equatorial wave appears to be excited between the middle and upper cloud (60-65 km altitude) and it propagates downward as a Kelvin wave, faster than zonal wind but also upward. At the cloud top, this wave propagates slower than the mean zonal wind, while recent Akatsuki UV observations suggest that the Kelvin wave propagates faster \citep{Imai2019}.
Kelvin waves with wavenumber-1 have also been predicted by other model experiments \citep[e.g][]{Yamamoto2006, Yamamoto2019, Sugimoto2014} but contrarily to our simulations they are slower near the cloud base ($\sim$ 50 km), with a period of 5-6 days.
On average, this perturbation shifts the convergence of equatorial zonal wind between LT 0h and 3h, as shown with the Venus-day averaged view of Figures~\ref{3D_map_CO_u}, or the instantaneous views of Figure~\ref{2D_maps_CO_TN}. In the second column of Figure~\ref{2D_maps_CO_TN} we show how CO, with a longer chemical lifetime
% of 10-100 days \citep{Brecht2011}, 
than the Kelvin wave period, accumulates preferentially towards the morning terminator, particularly at altitudes 90 to 110~km. On the contrary, atomic oxygen (not shown here), with a shorter life span, peaks near the AS point.
%Moreover, the CO bulge varies with the longitudinal phase of the Kelvin-like wave at those altitudes, with a more pronounced bulge during a Kelvin wave passes, creating a local maximum of CO concentration at 90 km, and a local minimum at 110 km at the morning terminator during half of the Kelvin wave period.
To our knowledge, there is no such observation of CO matching this GCM result.
Kelvin wave impact on the nightside also enhances the poleward meridional circulation. As a consequence, CO and O are also periodically transported to high latitudes likewise, as explained in Navarro2021 and shown in the third column of Figure~\ref{2D_maps_CO_TN}.
The poleward episodes create the high-latitude events of oxygen nightglow observed by VEx as high as 80$^\circ$ \citep{Soret2014}.
However, our GCM simulations did not produce such events at latitudes higher than 60$^\circ$, probably due to SS-AS too strong in the simulations.
Therefore, we would expect that high-latitude nighttime CO is also underestimated in these simulations. Figures~\ref{CO_SOIR_MT}-\ref{CO_SOIR_ET} instead show that the GCM largely overestimate SOIR measurements at the terminator in the latitude range 80$^\circ$N-90$^\circ$N. There are not CO nighttime measurements available at high-latitudes to contrast this result so far.
Another remarkable characteristic of the Venus upper atmosphere emerging from the IPSL-VGCM results is the quiet daytime upper mesosphere (above 100~km) compared with highly variable nighttime, in agreement with observations.

\subsection{Impact of non-orographic GW}
\label{GWimpact}
\begin{figure}[!htbp]
%\centering
\includegraphics[width=1.2\linewidth]{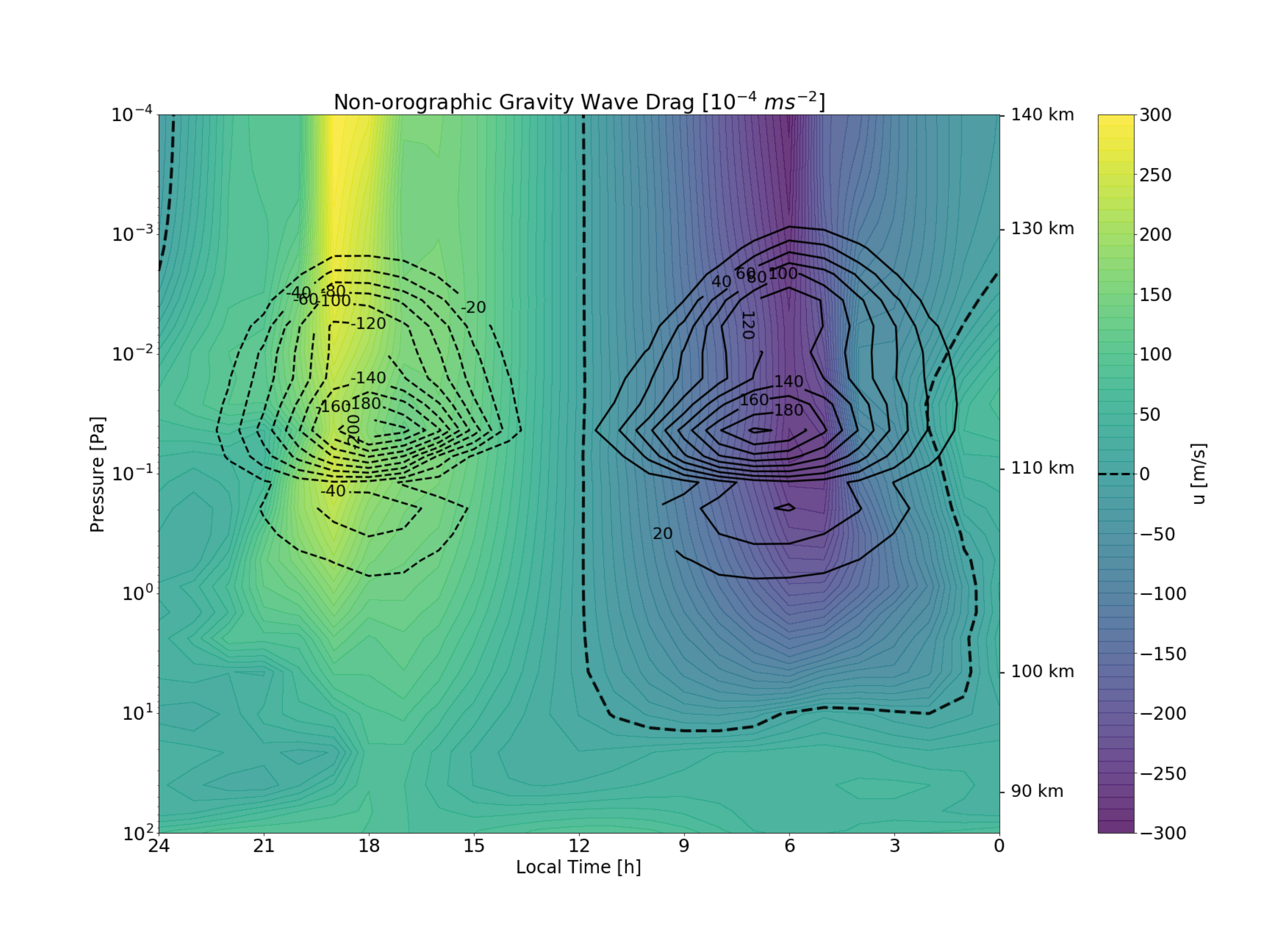}
%\subfloat[]{\includegraphics[width=0.7\linewidth]{Figures2/Non-orographicGW_equator_HR_day29_all_lats_normalised.pdf}}
%\qqauad
%\subfloat[]{\includegraphics[width=0.3\linewidth, height=7.3cm]{Figures2/nonGWudrag_profile2.pdf}}
%\vspace{1.cm}    
    \caption{Local time-pressure distribution of the GW zonal drag (contours) integrated over all latitudes ($\theta$) and normalized by $\cos$($\theta$) to account for conservation of angular momentum. Zonal winds (color) at equator regions (10$^\circ$S-10$^\circ$N) are also shown. Solid line indicates acceleration, dotted line deceleration. Positive zonal wind values are westward and negative zonal wind values are eastward. Dashed-line indicates the region where the zonal wind converges to zero. }
    \label{GWdrag_2D_map}
\end{figure}

%\textit{Impact of non-orographic GW:} 
Non-orographic GW breaking above 100 km also have a modest impact on the CO concentration. Previous theoretical studies (e.g. \citet{Hoshino2013} and \citet{Zalucha2013}) included a GW-drag parameterization in their model to investigate the effect of GW on the general circulation in the Venusian mesosphere and thermosphere. \citet{Hoshino2013} found the existence of weak wind layer at $\sim$125~km, not seen in previous simulations which used the Rayleigh friction scheme. Instead, \citet{Zalucha2013} suggested that gravity waves did not propagate above an altitude of nearly 115~km at the terminator because of total internal reflection. 
\\The non-orographic GW paramerization and the baseline parameters chosen here are described in \ref{Sec_inputs}. An analysis of the horizontal distribution of the wind velocity driven by GW is beyond the scope of this work, but the role of non-orographic GW forcing to the retrograde zonal flow in the transition region may be identified in our simulations.
They show a clear asymmetry of the SS-AS zonal flow between the morning and the evening branches and at altitudes above 100~km (see Figure~\ref{2D_maps_CO_TN}). 
As just explained in section \ref{KWimpact}, this asymmetry is caused by Kelvin wave impacting the mesosphere, but also by non-orographic GW. Indeed, these waves deposit momentum where they break, above 10$^{-1}$ Pa (110 km approximately), decelerating wind speed (e.g. causing acceleration of the SS-AS morning branch and deceleration of the SS-AS evening branch).
The GW drag predicted by our model is stronger at the terminator, as in previous studies \citep[e.g.][]{Alexander1992,Hoshino2013, Zalucha2013}, but has similar times values for the morning and evening terminator near the equator, as shown in Figure~\ref{GWdrag_2D_map}.
The deceleration of the evening branch of the SS-AS flow is stronger between LT 16h and LT 18h, with a maximum of $\sim$ 0.02 m s$^{-2}$ at 3$\times$10$^{-1}$ Pa. The acceleration instead occurs over larger local time and altitude ranges, reaching the maximum around LT 5h-8h (e.g. 0.018 m s$^{-2}$). 
Therefore, the mean zonal forcing integrated over local times (not shown here) is positive on average. All in all, the cause of the westwards dominant nightside winds and CO bulge towards the morning terminator is twofold: a westwards propagating Kelvin wave impacting the mesosphere, up to 110 km, and a weaker westwards acceleration by GW in the lower thermosphere, above 110 km.

\section{Conclusions}
\label{Conclusions}
This paper presents a comprehensive data-model comparison of temperature, CO$_2$, CO and O abundances in the Venus upper atmosphere using the results of an improved high resolution ground-to-thermosphere (0-150~km) version of the IPSL-VGCM. For this validation exercise we selected a collection of data from VEx experiments and coordinated ground-based campaigns available so far following Limaye17. 
\begin{itemize}
    \item 
Improvements of the non-LTE parameterization previously implemented in the IPSL-VGCM and described in \ref{nlte_improv} allow to predict a thermal structure in better agreement with observations. Both the intensity and the altitude of warm layer around 100~km is closer to measurements. The peak of temperature around midday and midnight is reduced by 40~K and 20~K, respectively and it is located about 10 km lower than in Gilli17. The temperature inversion observed by SOIR \citep{Vandaele2016} around 125~km, with a minimum of 125-130~K is also very well captured by our model.
Despite significant improvements, a number of discrepancies still remain. Nighttime temperature are still overestimated in the 90-115~km altitude region, by 20 to 30~K, with a peak temperature altitude $\sim$ 5 km higher than SPICAV \citep{Piccialli2015} retrieved profiles. Below 90~km, predicted temperatures tend to be at the upper limit of the range of observed values. A cold bias of 20~K to 50~K above 130~km is systematically found, compared to SOIR data. A candidate to explain this bias is the uncertainty in the collisional rate coefficients used in the non-LTE parameterization, that have an impact on the heating/cooling rate at thermosphere layers. 
%The sensitivity of the thermal structure of the current version of the IPSL-VGCM with this rate coefficient will be investigated in future GCM developments. \textbf{add this sentence somewhere in sec. 3}
Daytime simulated temperature profiles reproduce the warm and cold layers observed near 110 km and 125 km, respectively, but tend to overestimate their amplitude.   
%%% EXPLAIN HERE HOW TO IMPROVE THE REMAINING DISCREPANCIES, ex

\item
CO$_2$ observed densities are also in overall agreement with IPSL-VGCM results, with the exception of the equatorial region, where predicted values are 2-3 times larger than SPICAV  between 120 and 130 and than SOIR everywhere. Our model predicts very well density profiles measured during the VExADE aerobraking experiment \citep{Persson2015}.
The tion on the a-priori of CO$_2$ mixing ratio used to retrieve density and temperature in SOIR may partially explain the discrepancies that we found, in addition to the temperature cold bias found above 130 km at the terminator, mentioned above.

\item
Regarding the CO and O, model predictions are very well comparable with observations in order of magnitude and trend. Discrepancies found at daytime with VIRTIS data \citep{Gilli2015} at low/middle latitudes between 110 and 125 km (the model slightly underestimates the CO density) could be linked to the biases in the predicted temperature, the model being systematically warmer than the data between 90-110 km and colder between 120-150 km. At the terminator, CO densities are in overall agreement with SOIR dataset, in particular at the ET, but with larger differences (up to one order of magnitude between 90 and 100 km and a factor 2-3 elsewhere) at the MT, especially at low latitudes. %Simulated high-latitude (80$^\circ$N-90$^\circ$N) CO densities are also systematically higher than SOIR data.
O densities are in very good agreement with nighttime indirect measurements from O$_2$ nightglow, peaking at 2$\times 10^{11}$ cm$^{-3}$ around 110 km, as observed.

\item
The latitudinal and diurnal variation of CO and O has also been investigated. We found that the CO nighttime bulge is shifted toward the morning between 85 and 100~km, approximately, suggesting that CO is preferentially transported westwards in the transition region. This feature is supported by ground-based observations of CO at nighttime \citep{Gurwell1995,Lellouch2008,Moullet2012} which suggested the presence of a substantial westward retrograde zonal flow at altitudes above 90~km.
Atomic oxygen density instead peaks at midnight in the equatorial region  between 100 and 110 km, and they are about one order of magnitude larger than at noon, at the same altitude, as a result of efficient transport of atomic oxygen atoms from their day sources to their nighttime chemical loss at and below the peak.

\item
Our model includes a non-orographic GW parameterization following a stochastic approach as in \citet{Lott2012,Lott2013}. Both westward and eastward GW generated above typical convective cells propagate upwards. In our simulations they break mostly above 10$^{-1}$ Pa (110~km altitude approximately) and accelerate/decelerate zonal wind by momentum deposition near the terminator. The  mean zonal forcing is positive on average, contributing to retrograde imbalance of the flow in the lower thermosphere, above 110 km.
\\However, given the lack of systematic observations of GW, which are necessary to constrain model parameters, our experience with different GCM configurations let us conclude that the total zonal wind (i.e. averaged for all local times) value is very sensitive to many GCM quirks, and can be either negative or positive. Therefore, it is not excluded that this weak acceleration in the lower thermosphere is linked to the sensitivity of the model to unconstrained parameters.

\item
The increased horizontal resolution used in this paper had a negligible impact on the heating/cooling rate above 90 km compared to the lower resolution version in Gilli17, therefore a small influence on the validation of the temperature and densities. However, the shock-like feature above 110 km which reduces horizontal wind speeds and increases the number of large-scale eddies, is obtained only in the high-resolution case (see Navarro21). Those eddies may be responsible of the high variation observed in O$_2$ nightglow and not explained by other models using a lower resolution in the thermosphere \citep{Hoshino2013, Brecht2011}.

\item
%%%
A retrograde imbalance of the flow towards the morning side is predicted by our model at equatorial regions, and interpreted as a combination of non-orographic GW asymmetric drag on the zonal wind and perturbation produced by a Kelvin wave generated at the cloud deck. The impact of this $\sim$5 Earth day wave on the middle/upper atmosphere of Venus is described in details in the companion paper Navarro21.

%\item
%\textcolor{red}{
%Suggestion by Thomas, TBD: add recommendations about the needed observations to constrain GCMs: GW observations, mapping CO at scales that will let us test the possible influence of a Kelvin wave, etc \dots}

\end{itemize}

\section*{Acknowledgements}
GG was supported by the European Union's Horizon2020 research and innovation programme under the Marie Sklodowska-Curie grant agreement No. 796923 and by Fundaç\~ao para a Ci\^encia e a Tecnologia (FCT) through the research grants UIDB/04434/2020, UIDP/04434/2020, P-TUGA PTDC/FIS-AST/29942/2017.
The authors thank A.C. Vandaele for providing VAST and CO database from SOIR, and T. Clancy for sharing JCMT CO measurements.
TN and GS acknowledge NASA Akatsuki Participating Scientist Program under grant NNX16AC84G.
SL thanks the support from the Centre National d'Etudes Spatiales (CNES). FL thanks the Programme National de Planétologie (PNP) for financial support. This work used computational and storage services associated with the Hoffman2 Shared Cluster provided by UCLA Institute for Digital Research and Education’s Research Technology Group, as well as the High-Performance Computing (HPC) resources of Centre Informatique National de l’Enseignement Supérieur (CINES) under the allocations A0060110391 and A0080110391 made by Grand Equipment National de Calcul Intensif (GENCI).\\

\section*{Figures}

%%%%%%%%%%%%%%%%%%%%%%%%%%%%%%%%%%%%%%%%%%%%%%%%%%%%%%%%%%%%%%%%%%%%%%%%%
%
%      FIGURES
%
%%%%%%%%%%%%%%%%%%%%%%%%%%%%%%%%%%%%%%%%%%%%%%%%%%%%%%%%%%%%%%%%%%%%%%%%%%

%%%%%%%%%%%%%%%% %%%%%%%%%%%%%%%%%%%%%%%%%%%%%%%%%%%%%%%%%%%%%%%%%%%%%%%%%%%%
% FIGURE 1 TEMP PROFILES WITH SOIR/VEX improvement compared with GIlli2017
%%%%%%%%%%%%%%%%%%%%%%%%%%%%%%%%%%%%%%%%%%%%%%%%%%%%%%%%%%%%%%%%%%%%%%%%%%%%%

{}

%%%%%%%%%%%%%%%%%%%%%%%%%%%%%%%%%%%%%%%%%%%%%%%%%%%%%%%%%%%%%%%%%%%%%%%
% FIGURE 2 TEMP maps, LCT vs Pressure. Gilli2017 vs, best fit
%%%%%%%%%%%%%%%%%%%%%%%%%%%%%%%%%%%%%%%%%%%%%%%%%%%%%%%%%%%%%%%%%%%%%%

{}

%%%%% FIG 3

{}

%%%%% FIG 5
{}

%%%%% FIG 5
{}

%%%%%. FIG 7
{}
%%%%%%%%%%%%%%%%%%%%%%%%%%%%%%%%%%%%%%%%%%%%%%%%%%%%%%%%%%%%%%%%%%%%%%%%%%%%
% FIGURE 8
%  SOIR CO comparison, ET and MT
%   
%%%%%%%%%%%%%%%%%%%%%%%%%%%%%%%%%%%%%%%%%%%%%%%%%%%%%%%%%%%%%%%%%%%%%%%%%%%%%

{}

%%%%%. FIG 9
{}

%%%%%. FIG 10
{}

%%%%%. FIG 11
{}

%%%%%. FIG 12
{}

%&&&&&&&&&&&&&&&&&&&&&&&&&&&&&&&&&&&&&&&&&&&&&&&&&&&&&&&&&&&&&&&&&&
%%% Figures  Latitudinal variation of CO, and O VMR,  day and night
%%%%%%%%%%%%%%%%%%%%%

{}

{}

%&&&&&&&&&&&&&&&&&&&&&&&&&&&&&&&&&&&&&&&&&&&&&&&&&&&&&&&&&&&&&&&&&&
%%% Figures  14  CO Density: Diurnal variatio 
%%%%%%%%%%%%%%%%%%%%%

{}

{}

\clearpage

{}

%Panel\textit{(b)}: Local time integrated GW drag profile as function of pressure

\newpage

\appendix

\section{Non-orographic GW parameterization}
\label{GWparam}
\subsection{Formalism}
%Many different parameterizations including those of \cite{Lindzen1981} and \cite{Hines1997} have been used to represent the impact of subgrid scale gravity processes in the terrestrial atmosphere.
The scheme implemented here is based on a stochastic approach, as fully described  in \citet{Lott2012} and \citet{Lott2013}, also successfully implemented in the LMD Mars GCM \citep{Gilli2020}. In such a scheme, a finite number of waves (here M~=~8), with characteristics chosen randomly, are launched upwards at each physical time-step ($\delta t$ = 1.5 min) from every horizontal grid point to simulate their global effect. The question of the location of the non-orographic GW sources is a difficult one and given the lack of information on the location of GW sources, launching the gravity waves at each grid point, and at a fixed altitude, is probably the simplest assumption. The number of waves is given as M~=~NK~$\times$~NO~$\times$~NP, with NK~=~2 values of GW horizontal wave-numbers, NO~=~2 absolute values of phase speed, and NP~=~2 directions (westward and eastward) of phase speed.
This approach allows the model to treat a large number of waves at a given time $t$ by adding the effect of those M waves to that of the waves launched at previous steps, with a remnant time constant $\Delta t$ chosen to 1 Earth day (as for Earth and Mars).
%We need to parameterize GWs whose life cycle (i.e. from its generation to wave break) is contained in a characteristic time interval  $\Delta t$ that has to significantly exceed the GCM time step $\delta t$. On the Earth,  GW theory indicates that atmospheric disturbances induced by convection have life cycles with duration $\Delta t$ around 1 day ($\Delta t$ = 24h)\citep{Lott2013}. The choice for this timescale on Earth is made considering a mix of inertio-gravity waves (influenced by the planetary rotation, which is similar on Mars and on the Earth) and convectively-generated gravity waves, which frequency would be shorter than the inertio-gravity waves, but which propagation can occur over several hours before dissipation or breaking occurs.
\\The GW spectrum is then discretised in $\sim$ 7700 stochastic harmonics (M $\times \Delta t/ \delta t$) which contribute to the wave field each day and at a given horizontal grid point. $\Delta t$ is a characteristic time interval which contains the GW life cycle (i.e. from its generation to wave break) and which has to exceed significantly the GCM time step $\delta t$. On Earth,  GW theory indicates that atmospheric disturbances induced by convection have life cycles with duration $\Delta t$ around 1 day ($\Delta t$ = 24h)\citep{Lott2013}. Here, as a first approximation we adopted this value ($\Delta t \gg \delta t = 1.5$ min ).
At each time $t$ the vertical velocity field $w'$ of an upward propagating GW can be represented by the sum over harmonics:
\begin{equation}
	w' =\sum_{n=1}^{\infty} C_{n}  w'_{n}
\end{equation}  where $C_{n}$  are normalization coefficients such that $ \sum_{n=1}^{\infty} C_{n}^{2}  = 1$. It is assumed that each of the $w'_{n}$ can be treated independently from the others, and each $C_{n}^{2}$ can be viewed as the probability that the wave field is represented by $w'_{n}$ (see also the expression for $C_{n}$ in Equation \ref{Cn_equation}). This formalism is then applied to a very simple multi-wave parameterization, in which $w'_{n}$ represents a monochromatic wave as follows
\begin{equation}
\label{wn}
	w'_{n} =  \Re  \left\lbrace \hat{w}_{n}(z)e^{z/2H}e^{i(k_{n}x+l_{n}y- \omega_{n}t)}  \right\rbrace 
\end{equation}

where the wavenumbers $k_{n}$, $l_{n}$ and frequency $\omega_{n}$ are chosen randomly. In Equation \ref{wn}, H$\sim$4 km is a middle atmosphere characteristic vertical scale for Venus \citep{Lee2012, Mahieux2015} and z is the log-pressure altitude z = H $\ln$(P$_r$/P), with $P_r$ a reference pressure (P$_r$ = $5 \times 10^{4}$ Pa, taken here at the height of the source (i.e. above typical convective cells, $ z \sim$ 55 km). To evaluate the amplitude $\hat{w}_{n}(z)$, it is randomly chosen at a given launching altitude $z_0$, and then iterated from one model level $z_1$ to the next $z_2$ by a Wentzel-Kramers-Brillouin (WKB) approximation (see Equation 4 of \citet{Lott2012} for details). Using that expression plus the polarization relation between the amplitudes of large scale horizontal wind $\hat{u}$ and vertical wind $\hat{w}$ (not shown here), we can deduce the Eliassen-Palm (EP) flux (vertical momentum flux of waves),
\begin{equation}
\label{EPflux}
\vec{F}^z (k,l, \omega) = \Re \{ \rho_{r} \vec{\hat{u}} \hat{w^*} \}   = \rho_{r} \frac{\vec{k}}{| \vec{k}|^2} m(z) ||\hat{w}(z)||^2
\end{equation}
with $k,l$ the horizontal wavenumber and $\omega$ the frequency of the vertical velocity field. This last one is included in the WKB non-rotating approximation for the vertical wavenumber $m = \frac{N| \vec{k}|}{\Omega} $, with $\Omega = \omega - \vec{k} \vec{u} $ and N the Brunt-Vaisala frequency (see \citet{Lott2012} for details). 
In Equation (\ref{EPflux}), $\rho_{r}$ is the density at the reference pressure level $P_{r}$.
\\In this scheme, both $\hat{w}_{n}$ and the EP-flux at a given launching altitude $z_0$ (see Section \ref{Sec_inputs}) are randomly chosen. To move from one model level to the next model level above, the EP-flux is essentially conserved, but a small diffusivity is allowed, $\nu = \mu / \rho_0$, which can be included by replacing  $\Omega$ with $\Omega + i \nu m^2$. This small diffusivity is here to guarantee that the waves are ultimately dissipated over the few last model levels, if they have not been before (hence the division by the density $\rho_0$). In addition, this new EP-flux amplitude is limited to that produced by a saturated monochromatic wave $\hat{w}_s$ following \citet{Lindzen1981}:
\begin{equation}
\label{w_sat}
\hat{w}_s = S_c \frac{\Omega^2}{|\vec{k}| N} e^{-z/2H} \frac{k^*}{|\vec{k}|}	 
\end{equation}
or either $\hat{w}$ = 0 when $\Omega$ changes sign, to treat critical levels.
In Equation \ref{w_sat}, $S_c$ is a tunable parameter and $k^*$ a characteristic horizontal wavelength corresponding to the longest wave being parameterized (see more details in Sec. \ref{Sec_inputs})  
\\Finally, the accelerations $\rho^{-1} \delta_z \vec{F}^{z}_{n'}$   ($n'$= 1, M) produced by the GW drag on the winds are computed. Since $w'_n$s are independent realizations, the mean acceleration they produce is the average of these M acceleration. Thus, the averaged acceleration is first redistributed over the longer time scale $\Delta t$ by re-scaling it by $\delta t$/$\Delta t$ and second, the auto-regressive (AR-1) relation described in \citep{Lott2012} is used as follows:

\begin{equation}
\label{ARrelation}
  \left( \frac {\delta \vec{u}}{\delta t} \right)_{GWs}^{t} = \frac{\delta t}{\Delta t} \frac{1}{M} \sum_{n'=1}^{M} \frac{1}{\rho_0} \frac{\delta \vec{F}^{z}_{n'} }{\delta z} +  \frac{\Delta t - \delta t}{\Delta t}  \left( \frac {\delta \vec{u}}{\delta t} \right)_{GWs}^{t- \delta t} 
\end{equation}
This indicates that, at each time step, M new waves are promoted by giving them the largest probability to represent the GW field, and the probabilities of all the others are degraded by the multiplicative factor $(\Delta t - \delta t )/\Delta t$. As explained in \citet{Lott2012}, by expressing the cumulative sum underneath the AR-1 relation in Equation \ref{ARrelation}, the formalism for infinite superposition of stochastic waves is recovered by taking: 
\begin{equation}
\label{Cn_equation}
 C^{2}_{n} = \left( \frac{\Delta t - \delta t}{\Delta t}\right)^p \frac {\delta t}{M \Delta t}
\end{equation}
where $p$ is the nearest integer that rounds $(n-1)/M$ (i.e. toward lower values).

\subsection{GW parameters setup}
\label{Sec_inputs}
Here we describe the main tunable parameters used in the non-orographic GW scheme implemented in the IPSL-VGCM.  The characteristics of every wave launched in the GCM are selected randomly with a prescribed box-shaped probability distribution, whose boundaries are key model parameters. These are chosen on the basis of observational constraints (whenever available) and theoretical considerations: each parameter has a physical meaning, as described below.
\begin{description}
\item [Source height and duration]  
First, considering that non-orographic GW are expected to be generated by deep cloud convective layer on Venus, and mostly driven by longwave radiation, we assumed that the non-orographic GW source is placed above typical convective cells (i.e. at $5.\times 10^4$ Pa, around 55 km of altitude), turned on all day, uniform and without latitudinal variation.

 \item [EP-flux amplitude]  
$F^z$ from  Equation \ref{EPflux} gives the vertical rate of transfer of wave horizontal momentum  per units of area. This value has never been measured in the Venus atmosphere and it represents an important degree of freedom in the parameterization of gravity waves. Thus, in our scheme we impose the maximum value of the probability distribution  $F^{z}_{max}$ at the launching altitude z$_0$ (see previous point), at the beginning of the first run. 
In order to define $F^{0}_{max}$ we have explored typical values used in the literature to evaluate the order of magnitude of the EP-flux within the realm of what is realistic. 
It should be stressed here that the stochastic approach implemented in our scheme has the advantage of allowing to treat a wide diversity of emitted gravity waves, thereby a wide diversity of momentum fluxes. Our only setting is the maximum EP-flux amplitude at the launching altitude, that was fixed as 0.005 [kg m$^{-1}$ s$^{-2}$] (i.e. 5 mPa), following \citep{Lefevre2018} who simulated the gravity waves generated by the convective layer in a Large-Eddy-Simulation model describing realistically the Venusian convective activity.

\item [Horizontal wavenumber]	 
In our scheme the horizontal wave number amplitude is defined as in \cite{Lott2012} $ k* < |k| < k_s $. The minimum value is $k* = 1/ \sqrt{ \Delta{x} \Delta {y}}$, where  $\Delta x$ and $\Delta y$ are comparable with the GCM horizontal grid. The maximum (saturated) value is $k_s < N^2/u$, $N$ being the Brunt-Vaisala frequency associated to the mean flow, and $u_0$ the mean zonal wind at the launching altitude. We kept the minimum and maximum values for the GW horizontal wavelength ($\lambda_h$ = $2\pi/ k$ ) 50 km and 500 km, respectively as in Gilli17. Those values are within the observed range of GW wavelengths  \citep{Peralta2008,Piccialli2010, Tellmann2009, Altieri2014,Garcia2009} 
However, in order to correctly implement this parameterization with the increased resolution ($\delta$x and $\delta$y at the equator $\approx$ 400~km and 200 km in this study), we adapted the minimal horizontal wavenumber $k_{min}$ of individual waves as it follows:
\begin{equation}
    k^*_{min} = max(k_{min},\frac{\pi}{A}) 
\end{equation}
with $A$ the surface area of the grid cell.
This value corresponds to the longest waves that one parameterizes with the current  GCM horizontal resolution, larger wavelengths are explicitly resolved by the GCM.

\item [Phase speed] Another key  parameter is the amplitude of absolute phase speed $|c| = |\omega / k|$. As for the other tunable parameters, we impose the minimum $c_{min}$ and maximum  $c_{max}$ values of the probability distribution at the beginning of the runs, and the model chooses randomly $|c|$ between $c_{min}$ and  $c_{max}$.  Here $c_{min}$ is set to 1 m/s (i.e for non-stationary GW)  and $c_{max}$ is of the order of the zonal wind speed at the launching altitude ($u \approx 60~m/s$). Both eastward (c $>$ 0) and westward (c $<$ 0) moving GWs are considered.  

\item [Saturation parameter] $S_c$ is a tunable parameter in our scheme, on the right hand side of Equation \ref{w_sat}, which controls the breaking of the GW  by limiting the amplitude $w_s$. In the baseline simulation we set $S_c = 1$.
\end{description}

%%%%%%%%%%%%%%%%%%%%%%%%%%%%%%%%
% NLTE param improvements
%
%%%%%%%%%%%%%%%%%%%%%%%%%%%%%

\section{Improved Non-LTE parameterization set-up}
\label{nlte_improv}
The model includes 1-parameter formula that mimics the heating rates calculated by line-by-line non-LTE simulations for each pressure level $p$  and solar zenith angle $\mu$ (see Gilli17 for details). It is given by the following expression:

\begin{equation}
\frac{\partial T}{\partial t}(p, r, \mu) = \frac{\partial T}{\partial t}(p_0, r_0, 0) \times \frac{r_0^2}{r^2} \sqrt{\frac{p_0}{p} \widetilde{\mu}}\Big( 1 + \frac{p_1}{p} \Big) ^{-b}
\end{equation} 

\vspace{2mm}
where several orbital assumptions for Venus were considered: a circular planetary orbit with a mean solar distance of $r=r_0=$ 0.72 AU, no obliquity, and no seasonal variation. The cosine of the solar zenith angle $\widetilde{\mu}$ is corrected for atmospheric refraction using the following function: $\widetilde{\mu} = [(1224 \mu^2 + 1)/1225]^{\frac{1}{2}}$.

\vspace{2mm}
In this paper, the values of four non-LTE parameters used in Gilli17 are reviewed and adjusted: (i) cloud top pressure level in [$Pa$], $p_0$; (ii) heating rate per Venusian day (Vd) in [$K day^{-1}$], $\frac{\delta T}{\delta t}(p_0, r_0, 0)$; (iii) pressure below which non-LTE effects become significant in [$Pa$], $p_1$; and (iv) central pressure for transition from LTE to non-LTE radiation tendencies in [$Pa$], $ptrans$. 

\begin{figure}[!htbp]
    \centering
    \includegraphics[width=1.1\linewidth]{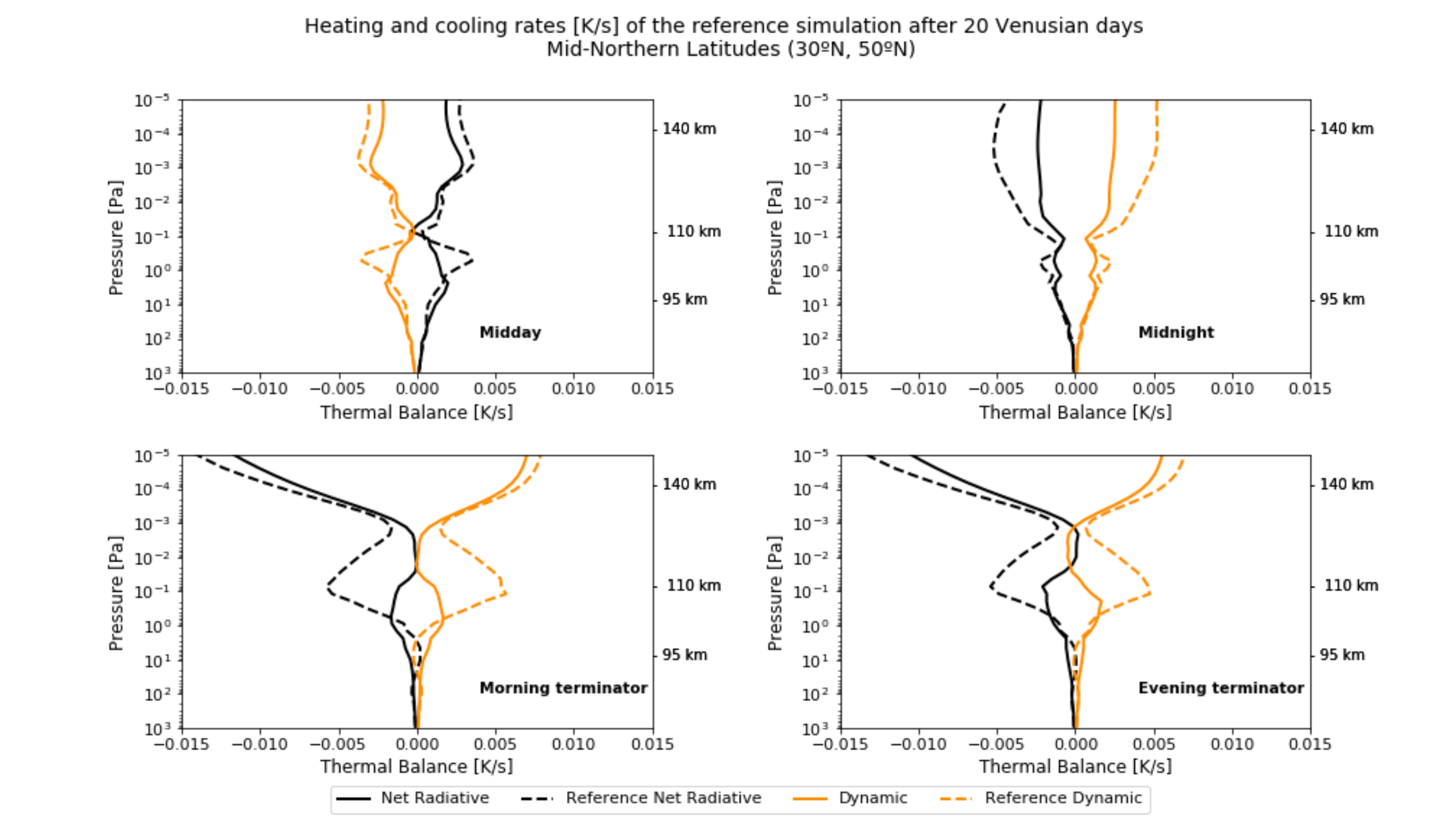}
    \caption{Net radiative and dynamical rates [K/s] obtained in this work for mid northern latitudes (30ºN-50ºN) after improving the Non-LTE parameterization for midday, nighttime and terminators, as indicated in the panels.  For comparison, the rates obtained in a previous version of the IPSL-VGCM (Gilli17), labeled as ``reference",  are also shown. Approximate altitudes are indicated on the right side of the panels. }
    \label{heating_Rates}
\end{figure}

First, we tuned the parameter $p_1$ to 0.1 [$Pa$] and $ptrans$ to 0.2 [$Pa$], which better represent the transition from LTE to non-LTE radiation. Second, following a literature review on the observed altitude of cloud tops \citep{Ignatiev2009, Haus2015} the mean cloud top altitude is approximately located between 69 and 74 km. We tested several pressure levels, $p_0$, in this range of altitudes with corresponding solar heating rates from look-up tables in \cite{Crisp1986}. The selected combination of those four parameters is listed in Table \ref{tableNLTE} together with those used in Gilli17, for comparison. 
%%%%%%%%%%%%%%%%%%%%%%%%%%%%%%%%
% FIGURE 1 HEATING/COOLING RATE.
%%%%%%%%%%%%%%%%%%%%%%%%%%%%%%%%
Figure \ref{heating_Rates} shows the predicted net radiative rates in the two works, at different local time.
The daytime heating/cooling peaking at 0.5 Pa (about 100 km) is reduced by almost a factor 2, while morning and evening terminator heating peaks around 0.1 Pa are about 3 times smaller, compared with Gilli17 simulations. Interestingly, nighttime radiative tendencies are affected by the changes of non-LTE parameter values mostly in the thermosphere, above 0.1 Pa (110 km approximately), suggesting that the region below is dominated by dynamical tendencies (e.g the adiabatic heating produces the local warm layer at nighttime), consistent with GCM results by \citet{Brecht2012a}.

%%%% FIG1

{}

\section{Bibliography styles}

%Here are two sample references: \cite{Feynman1963118,Dirac1953888}.

%\section*{References}
\newpage
\bibliography{Bibliografia_gab.bib}

\end{document}